\documentclass[aps,preprintnumbers,nofootinbib,floatfix]{revtex4}
\usepackage{amssymb,bm,graphicx}
\usepackage{amsmath,bbold,ulem}
\usepackage{dsfont}
\usepackage{slashed}
\usepackage[dvipsnames]{xcolor}
\usepackage[utf8]{inputenc}

\newcommand{\be}{\begin{equation}}
\newcommand{\ee}{\end{equation}}
\newcommand{\ba}{\begin{eqnarray}}
\newcommand{\ea}{\end{eqnarray}}
\newcommand{\bsub}{\begin{subequations}}
\newcommand{\esub}{\end{subequations}}
\newcommand{\la}{\langle}
\newcommand{\ra}{\rangle}
\newcommand{\ds}[1]{\slashed{#1}}

\newcommand{\blue}[1]{{\color{blue} #1}}
\newcommand{\red}[1]{{\color{red} #1}}

\begin{document}

\title{Structure of the nucleon at leading and subleading twist in the covariant parton model}

\author{
S.~Bastami$^1$, 
A.~V.~Efremov$^2$, 
P.~Schweitzer$^1$, 
O.~V.~Teryaev$^2$, 
P.~Zavada$^3$}
  \affiliation{
  $^1$ Department of Physics, University of Connecticut, 
  Storrs, CT 06269, U.S.A.\\
  $^2$ Bogoliubov Laboratory of Theoretical Physics, 
  JINR, 141980 Dubna, Russia\\
  $^3$ 
  Institute of Physics of the Czech Academy of Sciences, 
  Na Slovance 2, 182 21 Prague, Czech Republic}

\begin{abstract}
The covariant parton model is generalized to describe quark correlators
in a systematic way. Previous results are reproduced for the T-even 
leading-twist transverse momentum dependent parton distribution
functions (TMDs), and for the first time all T-even twist-3 TMDs are 
evaluated in this model. We apply the approach to evaluate the 
fully unintegrated quark correlator which allows us to understand
the model-specific relations between different TMDs. 
We verify the consistency of the approach, present numerical results
and compare to available TMD parametrizations.
\end{abstract}

\maketitle

\section{Introduction}
\label{intro}

TMDs allow one to explore the transverse structure of the nucleon in deeply 
inelastic scattering processes and are among the motivations for the planning 
of the Electron-Ion Collider \cite{Accardi:2012qut}. They enter the description 
of processes like semi-inclusive deep-inelastic scattering (SIDIS) 
\cite{Kotzinian:1994dv,Mulders:1995dh,Bacchetta:2006tn} 
(in conjunction with fragmentation functions \cite{Metz:2016swz}) 
or Drell-Yan \cite{Boer:1997nt,Arnold:2008kf} 
on the basis of TMD factorization theorems 
\cite{Collins:1981uk,Efremov:1981sh,Efremov:1984ip,Collins:1984kg,Qiu:1991pp,Ji:2004wu,Ji:2006br,Ji:2006vf,Collins:2011zzd,Aybat:2011zv,Bacchetta:2013pqa,Sun:2013dya,Echevarria:2014xaa,Collins:2014jpa,Collins:2016hqq}.
There has been an impressive progress in higher order QCD calculations
\cite{Gehrmann:2014yya,Echevarria:2015byo,Echevarria:2016scs,Li:2016ctv,Vladimirov:2016dll,Gutierrez-Reyes:2017glx,Gutierrez-Reyes:2018iod,Luo:2019hmp,Luo:2019szz,Ebert:2020yqt} and phenomenological studies
\cite{Efremov:2004tp,Anselmino:2005nn,Vogelsang:2005cs,Collins:2005ie,Collins:2005rq,Anselmino:2007fs,Anselmino:2013vqa,Signori:2013mda,Anselmino:2013lza,Kang:2014zza,Kang:2015msa,Kang:2017btw,Cammarota:2020qcw,Lefky:2014eia}.
Aspects of TMD physics were reviewed in Refs.~%
\cite{Collins:2003fm,DAlesio:2007bjf,Barone:2010zz,Aidala:2012mv,Avakian:2019drf,Anselmino:2020vlp}.
An important complement of the theoretical and phenomenological studies
is provided by studies in models. 

Models seek to explain the physics underlying phenomenological
observations in simple terms by focusing on certain concepts.
While often oversimplifying the complexities of hadronic physics,
models are nevertheless insightful, guide our intuition, and 
deepen the understanding of nucleon structure. One value of the
models is that they allow us to quantify how much of an observed
phenomenon can be attributed to a specific model concept. Another
important value is that models provide predictions for yet unknown 
nucleon properties which we can test in experiment.

This work deals with the study of TMDs in the covariant parton model 
(CPM) \cite{Zavada:1996kp,Zavada:2001bq,Zavada:2002uz,Efremov:2004tz,
Zavada:2007ww,Efremov:2009ze,Avakian:2009jt,Zavada:2009ska,
Efremov:2010mt,Zavada:2011cv,Zavada:2013ola,Zavada:2015gaa,Zavada:2019yom}
originally developed as a model for the description of the hadronic 
tensor in DIS \cite{Zavada:1996kp,Zavada:2001bq,Zavada:2002uz}. It is 
a parton model in the sense that the partons are free and non-interacting.
In the Feynman parton model this is the case in the infinite momentum
frame \cite{Feynman:1973xc}. In this sense the CPM goes one step further, 
and assumes the partons to be free in any frame. As free particles the 
partons are consequently on-shell. At the heart of the CPM are two 
types of covariant functions, $G^q(Pp)$ and $H^q(Pp)$, which describe 
the distributions of the momenta of respectively unpolarized and polarized 
partons inside the nucleon. The covariant distributions are functions of 
the scalar $Pp$ where $P$ is the nucleon and $p$ parton momentum.

The original formulation \cite{Zavada:1996kp,Zavada:2001bq,Zavada:2002uz}
allows one to evaluate the parton distribution functions (PDFs)
$f_1^a(x)$, $g_1^a(x)$, $g_T^a(x)$ accessible through DIS structure functions 
(throughout this work we do not indicate the scale dependence explicitly).
By an auxiliary polarized process due to the interference of vector and 
scalar currents, the approach was extended to the description of a 
hypothetical chiral-odd structure function and the transversity PDF 
$h_1^a(x)$ \cite{Efremov:2004tz}. The model was further generalized, 
by introducing the concept of ``unintegrated structure functions,'' 
to describe twist-2 T-even TMDs $f_1^a(x,p_T)$, $g_1^a(x,p_T)$, $h_1^a(x,p_T)$, 
$g_{1T}^{\perp a}(x,p_T)$, $h_{1L}^{\perp a}(x,p_T)$, $h_{1T}^{\perp a}(x,p_T)$
in \cite{Efremov:2009ze}. 
Despite these generalizations, the limitation is that many TMDs 
especially at twist-3 level cannot be studied in this way because no 
(real or auxiliary) process is known how to compute them in a 
parton model framework like the CMP.

The purpose of this work (after a brief review of quark correlators 
and TMDs in Sec.~\ref{Sec-2:correlator-TMDs})
is to generalize the formulation of the CPM to the description of quark 
correlators (in Sec.~\ref{Sec-3:fomulation-CPM}) which will put us
in the position to evaluate systematically all T-even TMDs.
We will reproduce earlier model results for twist-2 TMDs, and derive
new results for twist-3 TMDs (in Sec.~\ref{Sec-4:quark-TMDs}).
In order to test the internal theoretical consistency of the model
we will investigate the various emerging relations among TMDs,
some of which can be traced back to QCD equation-of-motion relations
or the so-called Lorentz-invariance relations which must be
valid in all quark models which respect Lorentz symmetry
(in Sec.~\ref{Sec-5:relations}). 
The relations among TMDs constitute one of the most interesting 
predictions of the CPM which can be tested quantitatively.
We will make predictions for all T-even unpolarized and polarized 
twist-2 and twist-3 TMDs (in Sec.~\ref{Sec-6:results}) and draw
conclusions (in Sec.~\ref{Sec-7:conclusions}). Technical details 
are presented in the Appendix. 

TMDs have been studied in bag 
\cite{Jaffe:1991ra,Yuan:2003wk,Courtoy:2008vi,Avakian:2008dz,
Courtoy:2008dn,Avakian:2010br}, 
quark-diquark 
\cite{Jakob:1997wg,Gamberg:2007wm,Cloet:2007em,Bacchetta:2008af,
She:2009jq,Lu:2012gu,Maji:2015vsa,Maji:2016yqo,Maji:2017bcz}, 
chiral quark soliton  
\cite{Diakonov:1996sr,Diakonov:1997vc,Gamberg:1998vg,Pobylitsa:1998tk,
Goeke:2000wv,Wakamatsu:2000fd,Schweitzer:2001sr,Schweitzer:2003uy,
Wakamatsu:2003uu,Ohnishi:2003mf,Cebulla:2007ej,Wakamatsu:2009fn,
Schweitzer:2012hh},
light-front constituent quark 
\cite{Pasquini:2008ax,Pasquini:2010af,Lorce:2011dv,
Boffi:2009sh,Pasquini:2011tk,Lorce:2014hxa,Kofler:2017uzq,Pasquini:2018oyz},
Nambu--Jona-Lasinio \cite{Matevosyan:2011vj}, Valon \cite{Yazdi:2014zaa},
holographic \cite{Maji:2017wwd,Lyubovitskij:2020otz} and
quark-target 
\cite{Kundu:2001pk,Meissner:2007rx,Mukherjee:2009uy,Mukherjee:2010iw,Xu:2019xhk}
models, and in some cases model-independently in lattice QCD computations 
\cite{Hagler:2009mb,Musch:2010ka,Musch:2011er,
Chen:2016utp,Alexandrou:2016jqi,Yoon:2017qzo,Orginos:2017kos,Joo:2019jct}.
We will compare to the results from other models. 
In view of the variety of the approaches, it is interesting that some 
nonperturbative properties of TMDs are supported across a broad class 
of different models \cite{Lorce:2011zta,Lorce:2011kn}.
The results presented in this work contribute to a picture of TMDs 
emerging from models, and help to solidify the understanding of TMDs.


\section{Quark Correlator and TMDs}
\label{Sec-2:correlator-TMDs}

The quark and antiquark correlation functions are defined as
\be\label{Eq:unint-correlator}
	\Phi_{ij}^q(p,P,S) = \int \frac{\mathrm{d}^4z}{(2\pi)^4} \; 
        \mathrm{e}^{i p z} \; \langle N(P,S) | \bar\psi_j^q(0) \; 
        \mathcal{W}(0,z;path) \; \psi_i^q(z)| N(P,S) \rangle\,,
\ee
where $p$ is the quark momentum. $P$ and $S$ denote the nucleon 
momentum and polarization with $P^2=M^2$, $S^2=-1$, $P\cdot S =0$. 
The correlator depends furthermore on a light-like 4-vector often 
denoted by $n^\mu$ which describes the lightcone direction. 
The Wilson line $\mathcal{W}(0,z;path)$, which is symbolically 
indicated in (\ref{Eq:unint-correlator}), depends on $n^\mu$ and 
the considered process. Depending on the chosen path the correlator 
may be relevant for DIS, Drell-Yan or another process
\cite{Collins:2002kn,Brodsky:2002rv,Brodsky:2002cx,Belitsky:2002sm,
Bomhof:2006dp}.
Strictly speaking we should denote the correlator as $\Phi_{ij}^q(p,P,S,n)$,
but for brevity we do not indicate the $n^\mu$-dependence (which is absent 
in the applications to quark models we have in mind).

Using lightcone coordinates, $p^\pm=\frac{1}{\sqrt{2}}(p^0\pm p^1)$,
one introduces the integrated correlators 
\be\label{Eq:int-correlator}
	\phi_{ij}^q(x,p_T, S) =
        \iint \mathrm{d} p^-\mathrm{d} p^+ \; \Phi_{ij}^q(p,P,S) 
        \; \; \delta(p^+-x\,P^+)\,,
\ee
If we define $\phi^{q[\Gamma]}=\frac12\,{\rm Tr}\,[\phi^q(x,p_T, S)\Gamma]$ 
then the leading twist quark TMDs are projected out from 
(\ref{Eq:int-correlator}) as follows
\begin{subequations}
\label{Eqs:def-TMDs}
\ba
    \phi^{q[\gamma^+]}
    &=& \hspace{5mm}
    \blue{f_1^q}-\frac{\varepsilon^{jk}p_T^j S_T^k}{M}\,\red{f_{1T}^{\perp q}}\;, 
    \label{Eq:TMD-pdfs-I}\\
    \phi^{q[\gamma^+ \gamma_5]} &=&
    S_L\,\blue{g_1^q} + \frac{\vec{p}_T\vec{S}_T}{M}\,\blue{g_{1T}^{\perp q}}\;, 
    \label{Eq:TMD-pdfs-II}\\
    \phi^{q[i \sigma^{j+} \gamma_5]} &=&
    S_T^j\,\blue{h_1^q}  + S_L\,\frac{p_T^j}{M}\,\blue{h_{1L}^{\perp q}} +
    \frac{\kappa^{jk}S_T^k}{M^2}\,
    \blue{h_{1T}^{\perp q}} + \frac{\varepsilon^{jk}p_T^k}{M}\,\red{h_1^{\perp q}}\;, 
    \label{Eq:TMD-pdfs-III} \hspace{15mm}
\ea
and the twist-3 quark TMDs are given by
\ba
\hspace{-5mm}    
	\phi^{q[\mathbb{1}]}        &=&
    	\frac{M}{P^+}\biggl[
	\;\blue{e^q}
	-\frac{\varepsilon^{jk}p_T^j S_T^k}{M}\,\red{e_T^{\perp q}}\,
    	\biggr], \label{Eq:sub-TMD-pdfs-I}\\
\hspace{-5mm}    
	\phi^{q[i \gamma^5]}       &=&
        \frac{M}{P^+}\biggl[
    	S_L \, \red{e_L^q} +\frac{\vec{p}_T \cdot \vec{S}_T}{M}\,\red{e_T^q}
    	\biggr], \label{Eq:sub-TMD-pdfs-II}\\
\hspace{-5mm}    
	\phi^{a[\gamma^j]}       &=&
        \frac{M}{P^+}\biggl[
    	\frac{p_T^j}{M}\blue{f^{\perp q}}\!+\varepsilon^{jk}S_T^k\red{f_T^q}
	\!+\!S_L\frac{\varepsilon^{jk}p_T^k}{M}\red{f_L^{\perp q}}
	\!-\!\frac{\kappa^{jk}\varepsilon^{kl}S_T^l}{M^2}\red{f_T^{\perp q}}\!
	\biggr], \label{Eq:sub-TMD-pdfs-III}\\
\hspace{-5mm}    
	\phi^{q[\gamma^j \gamma^5]} &=&
    	\frac{M}{P^+}\biggl[
    	S_T^j\,\blue{g_T^q} 
	+ S_L\,\frac{p_T^j}{M}\blue{g_L^{\perp q}} +
	\frac{\kappa^{jk}S_T^k}{M^2}
    	\,\blue{g_T^{\perp q}} 
	+\frac{\varepsilon^{jk}p_T^k}{M}\,\red{g^{\perp q}} 
	\biggr], \label{Eq:sub-TMD-pdfs-IV}\\
\hspace{-5mm}    
	\phi^{q[i \sigma^{jk} \gamma^5]} &=&
    	\frac{M}{P^+}\biggl[
    	\frac{S_T^j p_T^k-S_T^k p_T^j}{M}\,\blue{h_T^{\perp q}}
    	-\varepsilon^{jk}\,\red{h^q} 
	\biggr], \label{Eq:TMD-pdfs-V} \\
\hspace{-5mm}    
	\phi^{q[i \sigma^{+-} \gamma^5]} 
	&=& \frac{M}{P^+}\biggl[
    	S_L\,\blue{h_L^q} + \frac{\vec{p}_T\cdot\vec{S}_T}{M}\,\blue{h_T^q}
    	\biggr]. \label{Eq:TMD-pdfs-VI}
\ea\end{subequations}
The indices $j,\,k$ denote spatial directions transverse to the
lightcone, $\kappa^{jk}=(p_T^jp_T^k-\frac12\delta^{jk}p_T^2)$ where 
$p_T^2=|\vec{p}_T|^2$, and $\varepsilon^{23}=-\varepsilon^{32}=1$ and zero else.
In (\ref{Eqs:def-TMDs}) it is understood that $f_1^q=f^q_1(x,p_T)$, etc.
T-even TMDs (in blue color) can be computed in models based on
quark degrees of freedom only. T-odd TMDs (in red color)
require explicit gauge field degrees of freedom, 
and cannot be modeled in the approach used in this work.

The fully unintegrated quark correlator (\ref{Eq:unint-correlator}) 
has the following expansion in terms of Lorentz-invariant amplitudes 
\cite{Goeke:2005hb} 
\ba
\label{eq:corr-fully-unint}
     \Phi^q(P,p,S) \; &= \;& 
     MA_1^q + \slashed{P} A_2^q + \slashed{p}A_3^q 
     + \frac{\i}{2M} \; [\slashed{P},\slashed{p}] \; A_4^q 
     + i (p \cdot S) \gamma_5 \; A_5^q + M\slashed{S} \gamma_5 \; A_6^q
     \nonumber\\
     &&+ \frac{p \cdot S}{M} \slashed{P} \gamma_5 \; A_7^q + 
     \frac{p \cdot S}{M} \slashed{p} \gamma_5 \; A_8^q + 
     \frac{[\slashed{P},\slashed{S}]}{2} \gamma_5 \; A_9^q + 
     \frac{[\slashed{p},\slashed{S}]}{2} \gamma_5 \; A_{10}^q \nonumber \\
     &&+ \frac{p \cdot S}{2M^2} [\slashed{P},\slashed{p}] \gamma_5 \; A_{11}^q 
     + \frac{1}{M} \varepsilon^{\mu\nu\rho\sigma} \gamma_{\mu} P_{\nu} p_{\rho} 
     S_{\sigma} \; A_{12}^q + {\cal O}(B_i)
\ea
where $\varepsilon^{0123} = 1$. 
The amplitudes $A_4^q$, $A_5^q$, $A_{12}^q$ are T-odd. ${\cal O}(B_i)$ 
indicates symbolically the $B_1^q, \,\dots\,,\,B_{20}^q$ amplitudes 
associated with the lightlike vector $n^\mu$ inherent in the Wilson line. 
In models without gauge field degrees of freedom T-odd $A_i$ and all 
$B_i$ amplitudes are absent, and T-even TMDs are expressed in terms 
of the T-even $A_i^q$ as
\bsub
\label{Eq:TMDs-as-amp}
\begin{eqnarray}
         \label{eq:app_f_1}
         f_1^q(x,p_T) \; &=& \; 
         2P^+ \int \mathrm{d} p^- \biggl(A_2^q + x A_3^q\biggr) \, , \\
         \label{eq:app_f^perp_1T}
         g_1^q(x,p_T) \; &=& \; 
         2P^+ \int \mathrm{d} p^- 
         \biggl( -A_6^q-\frac{P \cdot p-M^2x}{M^2}(A_7^q+xA_8^q)\biggr) 
         + {\cal O}(B_i)\, , \\
         \label{eq:app_g_1T}
         g_{1T}^{\perp q}(x,p_T) \; &=& \; 
         2P^+ \int \mathrm{d} p^-\biggl(A_7+xA_8\biggr) \, , \\
         \label{eq:app_h_1}
         h_1^q(x,p_T) \; &=& \; 
         2P^+ \int \mathrm{d} p^- 
         \biggl( -A_9^q-xA_{10}^q+\frac{\vec{p}_T^{\,2}}{2M^2} \; A_{11}^q \biggr) 
         \, ,\\
         \label{eq:app_h^perp_1L}
         h^{\perp q}_{1L}(x,p_T) \; &=& \; 
         2P^+ \int \mathrm{d} p^- 
         \biggl( A_{10}^q - \frac{P \cdot p - M^2x}{M^2} \; A_{11}^q\biggr) 
         + {\cal O}(B_i)\, , \\
         \label{eq:app_h^perp_1T}
         h^{\perp q}_{1T}(x,p_T) \; &=& \; 
         2P^+ \int \mathrm{d} p^- A_{11}^q \, , \\
         \label{eq:app_e}
         e^q(x,p_T) \; &=& \; 
         2P^+ \int \mathrm{d} p^- A_1^q \, , \\
         \label{eq:app_f^perp}
         f^{\perp q}(x,p_T) \; &=& \; 
         2P^+ \int \mathrm{d} p^- A_3^q \, , \\
         \label{eq:appg_T}
         g_T^q(x,p_T) \; &=& \; 
         2P^+ \int \mathrm{d} p^- 
         \biggl( -A_6^q + \frac{\vec{p}_T^{\,2}}{2M^2} \; A_8^q \biggr) \, , \\
         \label{eq:app_g^perp_L}
         g^{\perp q}_L(x,p_T) \; &=& \; 
         2P^+ \int \mathrm{d} p^- 
         \biggl( - \frac{P \cdot p-M^2x}{M^2}\; A_8^q\biggr) 
         + {\cal O}(B_i)\, , \\
         \label{eq:app_g^perp_T}
         g^{\perp q}_T(x,p_T) \; &=& \; 
         2P^+ \int \mathrm{d} p^- A_8^q \, , \\
         \label{eq:app_h^perp_T}
         h^{\perp q}_T(x,p_T) \; &=& \; 
         2P^+ \int \mathrm{d} p^- \biggl(-A_{10}^q\biggr) \, , \\
         \label{eq:app_h_L}
         h_L^q(x,p_T) \; &=& \; 
         2P^+ \int \mathrm{d} p^- 
         \biggl( -A_9^q-\frac{P \cdot p}{M^2} \; A_{10}^q 
         + \biggl( \frac{P \cdot p-M^2x}{M^2} \biggr)^2 A_{11}^q \biggr) 
         + {\cal O}(B_i) \, , \\
         \label{eq:app_h_T}
         h_T^q(x,p_T) \; &=& \; 
         2P^+ \int \mathrm{d} p^- 
         \biggl( - \frac{P \cdot p-M^2x}{M^2} \; A_{11}^q\biggr) 
         + {\cal O}(B_i)\, .
\end{eqnarray}
\esub
We only symbolically indicate the $B_i^q$ amplitudes as they are absent 
in quark models, cf.~Ref.~\cite{Metz:2008ib} for the full expressions.
We also do not show the expressions for T-odd TMDs as they vanish
in quark models with no explicit gluon degrees of freedom
\cite{Pobylitsa:2003ty}.


\section{Formulation of the covariant parton model}
\label{Sec-3:fomulation-CPM}

We define the quark correlator in the CPM as follows
\be\label{eq:CPM1}
	\Phi^q(p,P,S)_{ij}  = 2P^0\,
        \Theta(p^0)\,\delta(p^2 - m^2)\;\bar{u}_j(p) u_i(p) \times
        \begin{cases}
        \,\mathcal{G}^q(pP) \; & \mbox{unpolarized partons,} \\
          \mathcal{H}^q(pP) \; & \mbox{polarized partons.} 
        \end{cases}
\ee
The prefactor $2P^0$ is due to the covariant normalization 
$\la N(P^\prime,S)|N(P,S)\ra =
2P^0\,(2\pi)^3\delta^{(3)}(\vec{P}^{\,\,\prime}-\vec{P}^{\,})$ 
of the nucleon states in Eq.~(\ref{Eq:unint-correlator}).
The onshell condition of the quarks in the CPM is implemented in 
terms of the Lorentz-invariant function $\Theta(p^0) \delta(p^2-m^2)$. 
$\mathcal{G}^q(pP)$ describes the covariant momentum distribution 
of unpolarized quarks of flavor $q=u,\,d,\,\dots$ inside the nucleon, while
$\mathcal{H}^q(pP)$ describes the covariant distribution of polarized quarks.

In the CPM the quarks are on-shell which allows us 
to evaluate the bispinor expressions as
\ba\label{eq:bispinors-q}  
          \bar{u}(p)\,\Gamma \, u(p)  &=&  {\rm Tr}\big[\frac{1}{2}  \, 
          (\ds{p}+m)\,\big(1+\gamma^5\,\ds{\omega}\big)\,\Gamma\big] \,,
\ea
Here $\omega^\mu$ is the quark polarization vector which satisfies
$\omega^2=-1$ and $p\cdot\omega = 0$ and can be expressed in the CPM
in terms of $p^\mu$, $P^\mu$, $S^\mu$ as follows \cite{Zavada:2001bq}
\be\label{eq:omega}
	\omega^\mu = - \frac{M}{m} \, \frac{p \cdot S}{p \cdot P + m M} 
        \, p^\mu - \frac{p \cdot S}{p \cdot P + m M} \, P^\mu + S^\mu.
\ee
A more general expression for $\omega^\mu$ was given in 
\cite{Zavada:2011cv} which coincides with (\ref{eq:omega}) for
massless quarks. In this work we will use Eq.~(\ref{eq:omega}) for 
$\omega^\mu$ and explore the more general representation for $\omega$
from \cite{Zavada:2011cv} elsewhere.

By exploring Eq.~(\ref{eq:bispinors-q}) we obtain the following compact 
expression
\ba\label{eq:CPM2}
        {\rm Tr}[\Phi^q(p,P,S)\,\Gamma]  &=& 
        P^0 \, \Theta(p^0)\,\delta(p^2 - m^2)\;
        {\rm Tr} \big[(\ds{p}+m)\,\bigl(\mathcal{G}^q(pP)
        +\mathcal{H}^q(pP)\gamma^5\,\ds{\omega}\bigr)\,\Gamma \big] \,.
\ea
We recall that the covariant function ${\cal G}^q(pP)$ is positive and has 
a partonic interpretation within the CPM: when interpreted in the nucleon 
rest frame it describes the momentum distribution of unpolarized quarks in
the nucleon. Similarly ${\cal H}^q(pP)$ describes the momentum distribution 
of polarized $q$ in a nucleon polarized in its rest frame along 
$S^\mu = (0,\vec{S})$ \cite{Zavada:1996kp,Zavada:2001bq,Zavada:2002uz}.
The unpolarized covariant function satisfies $G^q(Pp)\ge 0$, and the
polarized one $|H^q(Pp)|\le G^q(Pp)$ which reflects the partonic
interpretation.

The Eqs.~(\ref{eq:CPM1}--\ref{eq:CPM2}) can be viewed as a 
definition of the CPM and describe how to evaluate in the model 
quark correlation functions. 
In the remainder of this work, we will compute all twist-2 and 
twist-3 T-even TMDs of quarks on the basis of 
Eqs.~(\ref{eq:CPM1}--\ref{eq:CPM2}). Hereby we will
reproduce results known from previous works, derive many new 
results (especially for twist-3 TMDs), and demonstrate the internal 
theoretical consistency of the approach.


\section{Quark TMD\lowercase{s} in the covariant parton model}
\label{Sec-4:quark-TMDs}

This section is devoted to quark TMDs. In the twist-2 case we will 
rederive results obtained in Ref.~\cite{Efremov:2009ze} in different ways.
In the twist-3 case we will (with one exception) present new predictions.

\subsection{The unpolarized leading twist TMD \boldmath $f_1^q(x,p_T)$}
\label{Sec4a:twist-2-unpol}

In order to derive the expression for the TMD $f_1^q(x,p_T)$ we evaluate
the correlator
\ba\label{eq8}
	\phi^{q[\gamma^+]}(x,\vec{p}_T,S) 
          &=&
        \int \mathrm{d}p^- \mathrm{d}p^+ \; P^0 \, \mathcal{G}^q(pP) 
        \; \Theta(p^0) \, \delta(p^2 - m^2) \; \; \delta(p^+-x\,P^+)\,
        \; \bar{u}(p)\,\gamma^+ u(p) \,,\nonumber\\
          &=&
        \int \mathrm{d}p^- \mathrm{d}p^+ \; 
        \frac{ P^0  \, \mathcal{G}^q(pP)}{P^+}
        \; \Theta(p^0) \, \delta(p^2 - m^2) \; \; 
        \delta(x-\frac{p^+}{P^+}) 
        \; \bar{u}(p)\,\gamma^+ u(p) \,,
\ea
We choose to work in the nucleon rest frame where $pP=p^0M$. 
In the following we will denote the covariant function $\mathcal{G}^q(pP)$ 
in the nucleon rest frame for simplicity by $\mathcal{G}^q(p^0)$. 
In the nucleon rest frame $p^+/P^+=(p^0+p^1)/M$. To arrive at the 
formulation of the CPM from prior works we change the integration variables 
$\mathrm{d}p^- \mathrm{d}p^+ \to \mathrm{d}p^0 \mathrm{d}p^1$ such that 
\be\label{Eq:integrate-delta}
       \iint \mathrm{d}p^- \mathrm{d}p^+ 
        \; \Theta(p^0) \, \delta(p^2 - m^2) \;\dots =
       \iint \mathrm{d}p^0 \mathrm{d}p^1 
        \; \Theta(p^0) \, \delta\big((p^0)^2-\vec{p}^{\,2}- m^2\big) 
        \;\dots =
       \int \frac{\mathrm{d}p^1}{2p^0} 
        \; \dots\;\biggl|_{p^0=\sqrt{\vec{p}^{\,2}+m^2}} \,.
\ee
Evaluating the bispinor expression $\bar{u}(p)\,\gamma^\mu u(p)=2p^\mu$ 
we obtain
\be
        \phi^{q[\gamma^+]}(x,\vec{p}_T,S) = f_1^q(x,p_T) =
	\int \frac{\mathrm{d}p^1}{p^0}  \, \mathcal{G}^q(p^0)
          \; \delta(x-\frac{p^0+p^1}{M}) \; (p^0+p^1) \; . \label{Eq:f1}
\ee
This coincides with the result for $f_1^q(x,p_T)$ from Eq.~(25) 
in \cite{Efremov:2009ze} (after the substitution $p^1\to(-p^1)$).
In \cite{Efremov:2009ze} this result was obtained by introducing and
modeling a ``$p_T$-unintegrated'' hadronic tensor. Here we derive the
same result systematically from the model expression of the quark correlator 
(\ref{eq:CPM2}). 
We also see that in the model $\phi^{q[\gamma^+]}(x,\vec{p}_T,S)$ has no term 
proportional to $\vec{p}_T\vec{S}_T$ and hence the Sivers function is zero 
as expected in models with no gluons \cite{Pobylitsa:2003ty}.


\subsection{The chiral-even polarized leading twist TMD\lowercase{s}
\boldmath $g_1^q(x,p_T)$ and $g_{1T}^{\perp q}(x,p_T)$}
\label{Sec4b:twist-2-pol-chiral-even}

These TMDs require polarization. We use the expression (\ref{eq:omega}) 
for the quark polarization vector $\omega^\mu$ in nucleon rest frame 
where\footnote{%
   In Ref.~\cite{Efremov:2009ze} the lightcone spatial
   direction was chosen opposite to our work, i.e.\ the signs of the 
   first components of all vectors are reversed: for instance
   $(p^0-p^1)|_\text{\rm Ref.~\cite{Efremov:2009ze}}$ corresponds to 
   $(p^0+p^1)|_{\rm here}$. Consequently the conventions are such that
   $S_L|_{\rm here}=-S_L|_\text{\rm Ref.~\cite{Efremov:2009ze}}$.}
$S^\mu=(0,S_L,\vec{S}_T)$ in usual 4-vector notation. 
We explore (\ref{Eq:integrate-delta}) 
and $\bar{u}(p)\,\gamma^\mu\gamma_5 u(p)=2\omega^\mu$, and obtain
\begin{align} 
        \phi^{q[\gamma^+\gamma_5]}(x,\vec{p}_T,S) 
       =& \;S_L\;
        \int \frac{\mathrm{d}p^1}{p^0}  \, \mathcal{H}^q(p^0) 
        \; \delta(x-\frac{p^0+p^1}{M})\biggl[
           \frac{p^1(p^0+p^1)}{p^0 + m} 
        + \frac{p^1m}{p^0 + m}\
        + m\biggr] \nonumber\\
       +&\;\vec{p}_T\vec{S}_T\;
        \int \frac{\mathrm{d}p^1}{p^0}  \, \mathcal{H}^q(p^0) 
        \; \delta(x-\frac{p^0+p^1}{M}) \biggl[
          \frac{p^0+p^1}{p^0 + m} 
        + \frac{m}{p^0 + m}\,\biggr] \label{Eq:check-gamma+gamma5}
\end{align}
where we grouped terms proportional to longitudinal and transverse polarization.
Comparing to the coefficients in Eq.~(\ref{Eq:TMD-pdfs-II}) we read off the 
model results for $g_1^q(x,p_T)$ and $g_{1T}^{\perp q}(x,p_T)$ which, 
after some algebra, can be written as
\bsub\label{Eq:g1-g1Tperp}
\begin{align}
        g_1^q(x,p_T)
        =& \;
        \int \frac{\mathrm{d}p^1}{p^0}\,\mathcal{H}^q(p^0)\;
        \delta(x-\frac{p^0+p^1}{M}) 
        \,\biggl[p^0+p^1- \frac{p_T^2}{p^0 + m}\biggr]\\
        g_{1T}^{\perp q}(x,p_T)
        =& \;
        \int \frac{\mathrm{d}p^1}{p^0}\,\mathcal{H}^q(p^0)\;
        \delta(x-\frac{p^0+p^1}{M}) 
        \,\biggl[M\frac{p^0+p^1+m}{p^0 + m}\biggr] 
\end{align}\esub
and coincide with the expressions in Eqs.~(16,~17) of 
Ref.~\cite{Efremov:2009ze}. It is important to stress that these results
were obtained from the antisymmetric part of the ``$p_T$-unintegrated'' 
hadronic tensor which was constructed and modeled in \cite{Efremov:2009ze},
while here they follow straightforwardly from the model expression of the 
quark correlator (\ref{eq:CPM2}).


\subsection{The chiral-odd polarized leading twist TMD\lowercase{s}
\boldmath $h_1^q(x,p_T)$, $h_{1L}^{\perp q}(x,p_T)$, $h_{1T}^{\perp q}(x,p_T)$}
\label{Sec4c:twist-2-pol-chiral-odd}

In order to evaluate the correlator $\phi^{[i\sigma^{+j}]}$ 
we proceed as in (\ref{Eq:integrate-delta}), explore
$\bar{u}(p)\,i\sigma^{\mu\nu}u(p)=2\,(\omega^\mu p^\nu-\omega^\nu p^\mu)$
and insert the expression (\ref{eq:omega}) for $\omega^\mu$ with
$S^\mu=(0,S_L,\vec{S}_T)$. This yields
\begin{align} 
        \phi^{q[i\sigma^{+j}]}(x,\vec{p}_T,S) 
        &=
        \int \frac{\mathrm{d}p^1}{p^0}  \, \mathcal{H}^q(p^0) 
        \; \delta(x-\frac{p^0+p^1}{M}) \;
        \biggl((\omega^0+\omega^1) p^j-\omega^i (p^0+p^1)\biggr)\nonumber\\
        &=
        \int \frac{\mathrm{d}p^1}{p^0}  \, \mathcal{H}^q(p^0) 
        \; \delta(x-\frac{p^0+p^1}{M}) \;
        \biggl(
          S_T^j(p^0+p^1)
        - \frac{(p^0+p^1+m)}{(p^0+m)}\,S_L p_T^j
        - \frac{\vec{p}_T\vec{S}_T}{(p^0 + m)}\,p_T^j
        \biggr)
\end{align}
Notice that 
$(\vec{p}_T\cdot\vec{S}_T)\,p_T^j=\frac12\vec{p}_T^{\;2}S_T^j
+(p_T^jp_T^k-\frac12\delta^{jk}\vec{p}_T^{\;2})S_T^k$ where the
first term (monopole in $\vec{p}_T$) contributes to transversity
while the second term (quadrupole structure) gives rise to pretzelosity.
Comparing to the correlator (\ref{Eq:TMD-pdfs-III}) we read off 
the results
\bsub\label{Eq:TMD-twist2-chiral-odd}
\begin{align} 
        h_1^q(x,p_T) = &
        \int \frac{\mathrm{d}p^1}{p^0}  \, \mathcal{H}^q(p^0) 
        \; \delta(x-\frac{p^0+p^1}{M}) \;
        \biggl(p^0+p^1-\frac{p_T^2}{2(p^0+m)}\biggr) , \\
        h_{1L}^{\perp q}(x,p_T) = &
        \int \frac{\mathrm{d}p^1}{p^0}  \, \mathcal{H}^q(p^0) 
        \; \delta(x-\frac{p^0+p^1}{M}) \;
        \biggl(-M\;\frac{p^0+p^1+m}{p^0+m}\biggr) , \\
        h_{1T}^{\perp q}(x,p_T) = &
        \int \frac{\mathrm{d}p^1}{p^0}  \, \mathcal{H}^q(p^0) 
        \; \delta(x-\frac{p^0+p^1}{M}) \;
        \biggl(-\;\frac{M^2}{p^0+m}\biggr) .
\end{align}
\esub
As the model generates no unpolarized structure in the correlator 
(\ref{Eq:TMD-pdfs-III}), the T-odd Boer-Mulders function $h_1^{\perp q}(x,p_T)$ 
vanishes as expected in quark models \cite{Pobylitsa:2003ty}.
The results (\ref{Eq:TMD-twist2-chiral-odd}) agree with those obtained 
previously by generalizing the auxiliary polarized process due to 
interference of vector and scalar currents to the ``$p_T$-unintegrated'' 
situation \cite{Efremov:2009ze}.
Notice that the results can be simplified using e.g.\ $p^0+p^1=xM$ 
under the integrals in (\ref{Eq:TMD-twist2-chiral-odd}).

\subsection{Twist-3 TMD\lowercase{s}}
\label{Sec4d:twist-3}

In the twist-3 correlators (\ref{Eq:sub-TMD-pdfs-I}--\ref{Eq:sub-TMD-pdfs-IV}) 
we encounter 2 new Dirac structures, $\bar{u}(p)\,\mathds{1}\,u(p) = 2\,m$ 
related to $e^q(x,p_T)$ and $\bar{u}(p) \, i \gamma^5 \, u(p) =0$ 
for an onshell particle. Proceeding analog to the 
Secs.~\ref{Sec4a:twist-2-unpol}--\ref{Sec4c:twist-2-pol-chiral-odd} 
yields the following results
\bsub\label{eq:result-twist-3}
\begin{align}
  f^{\perp q}(x,p_T) \, = & \, \int \frac{\mathrm{d}p^1}{p^0}  \, 
 {\mathcal{H}^q(p^0)}\;\delta(x-\frac{p^0+p^1}{M})\,
  \frac{M (p^0+m)}{p^0+m} \label{Eq:model-fperp}\\
  e^q(x,p_T) \, = & \, \int \frac{\mathrm{d}p^1}{p^0}  \, 
 {\mathcal{H}^q(p^0)}\;\delta(x-\frac{p^0+p^1}{M})\, 
  \frac{m (p^0+m)}{p^0+m} \label{Eq:model-e}\\
  g_T^q(x,p_T) \, = & \, \int \frac{\mathrm{d}p^1}{p^0}  \, 
 {\mathcal{H}^q(p^0)}\;\delta(x-\frac{p^0+p^1}{M})\, 
  \frac{m (p^0+m) + \frac12\vec{p}_T^{\;2}}{p^0+m} \label{Eq:model-gT}\\
  g_T^{\perp q}(x,p_T) \, = & \, \int \frac{\mathrm{d}p^1}{p^0}  \, 
 {\mathcal{H}^q(p^0)}\;\delta(x-\frac{p^0+p^1}{M})\, 
  \frac{M^2 }{p^0+m}\label{Eq:model-gTperp}\\
  g_L^{\perp q}(x,p_T) \, = & \, \int \frac{\mathrm{d}p^1}{p^0}  \, 
 {\mathcal{H}^q(p^0)}\;\delta(x-\frac{p^0+p^1}{M})\, 
  \frac{M(Mx - p^0)}{p^0+m}  \label{Eq:model-gLperp}\\
  h_T^{\perp q}(x,p_T) \, = & \, \int \frac{\mathrm{d}p^1}{p^0}  \, 
 {\mathcal{H}^q(p^0)}\;\delta(x-\frac{p^0+p^1}{M})\,
  \frac{M (p^0+m)}{p^0+m} \label{Eq:model-hTperp}\\
  h_L^q(x,p_T) \, = & \, \int \frac{\mathrm{d}p^1}{p^0}  \, 
 {\mathcal{H}^q(p^0)}\;\delta(x-\frac{p^0+p^1}{M})\, 
  \frac{m (p^0+m) + \vec{p}_T^{\,2}}{p^0+m} \, \label{Eq:model-hL}\\
  h_T^q(x,p_T) \, = & \, \int \frac{\mathrm{d}p^1}{p^0}  \, 
 {\mathcal{H}^q(p^0)}\;\delta(x-\frac{p^0+p^1}{M})\, 
  \frac{M (p^0-M x)}{p^0+m} \,. \label{Eq:model-hT}
\end{align}
\esub
Only the model expression for the twist-3 TMD $g_T^q(x,p_T)$ was 
computed before in the CPM, as it is related to 
the (``unintegrated'' generalization of the) hadronic tensor in 
polarized DIS, and our result (\ref{Eq:model-gT}) agrees with 
the expression from \cite{Efremov:2009ze}.
The results for the other twist-3 TMDs are new. Notice that also 
in twist-3 case T-odd TMDs vanish in the CPM as it must be for a model 
with no explicit gauge field degrees of freedom \cite{Pobylitsa:2003ty}.

\section{Relations among TMDs}
\label{Sec-5:relations}

In QCD the different TMDs are all independent of each other, and 
describe different aspects of the nucleon structure. Due to the simpler
dynamics or additional symmetries, different TMDs can be related to each 
other in quark models. The goal of this section is to discuss the 
relations among TMDs in the CPM. 

\subsection{Equation-of-motion relations}
\label{sec:TMDrelations}

An important consistency test of the model results is provided by the 
following relations which can be derived by making use of the QCD 
equations of motion (EOM) and are given by
\bsub
\label{Eq:EOM}
\ba
  xe^q(x,p_T) 
  & = &  x\tilde{e}^q(x,p_T) + \frac{m}{M}\,f_1^q(x,p_T) , \label{Eq:EOM-e}\\
  xf^{\perp q}(x,p_T)   
  & = &  x\tilde{f}^{\perp q}(x,p_T) + f_{1}^q(x,p_T),
  \phantom{\frac11}\label{Eq:EOM-fperp}\\
  xg_L^{\perp q}(x,p_T) 
  & = &  x\tilde{g}_L^{\perp q}(x,p_T) + g_{1}^q(x,p_T) 
    + \frac{m}{M}\,h_{1L}^{\perp q}(x,p_T), \label{Eq:EOM-gLperp}\\
  xg_T^q(x,p_T)   	
  & = &  \tilde{g}_T^q(x,p_T) + g_{1T}^{\perp(1)q}(x,p_T)
    + \frac{m}{M}\,h_1^q(x,p_T), \label{Eq:EOM-gT}\\
  xg_T^{\perp q}(x,p_T)
  & = &  x\tilde{g}_T^{\perp q}(x,p_T) + g_{1T}^{\perp q}(x,p_T)
    + \frac{m}{M}\,h_{1T}^{\perp q}(x,p_T), \label{Eq:EOM-gTperp}\\
  xh_L^q(x,p_T)
  & = &  x\tilde{h}_L^q(x,p_T)-2\,h_{1L}^{\perp(1)q}(x,p_T) 
    + \frac{m}{M}\,g_1^q(x,p_T), \label{Eq:EOM-hL}\\
  xh_T^q(x,p_T)            
  & = &  x\tilde{h}_T^q(x,p_T) - h_1^q(x,p_T) - h_{1T}^{\perp(1)}(x,p_T)
    + \frac{m}{M}\,g_{1T}^\perp(x,p_T), \label{Eq:EOM-hT}\\
  xh_T^{\perp q}(x,p_T)     
  & = &  x\tilde{h}_T^{\perp q}(x,p_T) + h_1^q(x,p_T) - h_{1T}^{\perp(1)}(x,p_T),
    \phantom{\frac11} \label{Eq:EOM-hTperp}
\ea
\esub
where the transverse moment $n$ of a generic TMD $f^q(x,p_T)$ 
is defined as follows
\be\label{Eq:define-transv-mom}
       f^{(n)q}(x,p_T) = \biggl(\frac{p_T^2}{2M^2}\biggr)^{\!\!n}\,f^q(x,p_T).
\ee
The EOMs (\ref{Eq:EOM}) arise because the quark correlators defining 
twist-3 TMDs, Eqs.~(\ref{Eq:sub-TMD-pdfs-I}--\ref{Eq:sub-TMD-pdfs-IV}),
can be decomposed into contributions from quark-gluon correlators,
twist-2 correlators, and terms proportional to current quark masses 
by exploring QCD equations of motion \cite{Bacchetta:2006tn}.
The quark-gluon correlators give rise to ``genuine twist-3'' contributions 
or ``interaction dependent terms'' which are denoted by tilde-functions 
in (\ref{Eq:EOM}). In general the EOMs do not imply relations among TMDs, 
but {\it define} the respective genuine twist-3 tilde contributions.

In quark models in general the tilde terms are non-zero and arise from 
the model interactions due to the pertinent model equations of motion. 
The EOM of the CPM is the free Dirac equation which
implies the absence of tilde-terms. Our results for twist-2 and twist-3 TMDs,
Eqs.~(\ref{Eq:f1},~\ref{Eq:g1-g1Tperp},~\ref{Eq:TMD-twist2-chiral-odd},
\ref{eq:result-twist-3}), satisfy 
the EOM relations (\ref{Eq:EOM}) with the tilde-terms set to zero. 
This is an important consistency test for our new results 
for all twist-3 TMDs.

\subsection{Lorentz invariance relations}

In models with no gluonic degrees of freedom, such as the CPM, 
T-odd TMDs are absent \cite{Pobylitsa:2003ty} and the quark correlator can 
be decomposed in terms of 9 T-even $A_i$-amplitudes while it gives rise to 
14 T-even TMDs, see Sec.~\ref{Sec-2:correlator-TMDs}.  
As there are 14 TMDs and 9 linearly independent amplitudes, this implies 
5 relations among T-even TMDs. These relations are referred to as 
Lorentz-invariance relations (LIRs) and are not valid in QCD
\cite{Goeke:2003az} but must hold in all models which preserve Lorentz 
symmetry and exhibit no gauge degrees of freedom. The LIRs are given by 
\cite{Mulders:1995dh}
\bsub\label{Eq:LIRs}
\begin{align}
   \label{eq:LIR1} g_T^q(x) \; \stackrel{\rm LIR}{=}
   & \; g_1^q(x) + \frac{\mathrm{d}}{\mathrm{d} x} g^{\perp(1)q}_{1T}(x)\, ,\\ 
   \label{eq:LIR2} h_L^q(x) \; \stackrel{\rm LIR}{=}
   & \; h_1^q(x) - \frac{\mathrm{d}}{\mathrm{d} x} h^{\perp(1)q}_{1L}(x) \, , \\
   \label{eq:LIR3} h_T^q(x) \;  \stackrel{\rm LIR}{=}
   & \; - \frac{\mathrm{d}}{\mathrm{d} x} h^{\perp(1)q}_{1T}(x) \, , \\
   \label{eq:LIR4} g_L^{\perp q}(x) + \frac{\mathrm{d}}{\mathrm{d} x} g_T^{\perp(1)q}(x)   \;\stackrel{\rm LIR}{=}& \; 0 \, ,\\ 
   \label{eq:LIR5}
   h_T^q(x,p_T)-h_T^{\perp q}(x,p_T) \; \stackrel{\rm LIR}{=}
   & \; h^{\perp q}_{1L}(x,p_T) \, ,
\end{align}
\esub
and connect twist-3 TMDs (on left-hand sides of the above equations)
with twist-2 TMDs (if any, on right hand sides). The CPM satisfies all LIRs. 
This is an important consistency test for the model.
The Eqs.~(\ref{eq:LIR1}--\ref{eq:LIR4}) can be proven using
the methods developed in App.~C of Ref.~\cite{Efremov:2009ze},
while (\ref{eq:LIR5}) corresponds to the quark model
relation (\ref{Eq:qm-rel-6}) which is discussed below.

\subsection{Quark model relations}
\label{Sec-6:qm-relations}

Our results for the twist-2 and twist-3 TMDs satisfy also the 
following relations
\cite{Avakian:2010br}
\bsub\label{Eq:quark-model-relations}
\ba
      g_{1T}^{\perp q}(x,p_T) &=& - h_{1L}^{\perp q}(x,p_T), 
      \label{Eq:qm-rel-1}\\
      g_{ T}^{\perp q}(x,p_T) &=& - h_{1T}^{\perp q}(x,p_T), 
      \label{Eq:qm-rel-2}\\
      g_{ L}^{\perp q}(x,p_T) &=& - h_{ T}^{q}(x,p_T), 
      \label{Eq:qm-rel-3}\\
      g_1^q(x,p_T)-h_1^q(x,p_T) &=& h_{1T}^{\perp(1)q}(x,p_T), 
      \label{Eq:qm-rel-4}\\
      g_T^q(x,p_T)-h_L^q(x,p_T) &=& h_{1T}^{\perp(1)q}(x,p_T), 
      \label{Eq:qm-rel-5}\\
      h_T^q(x,p_T)-h_T^{\perp q}(x,p_T) &=& h_{1L}^{\perp q}(x,p_T). 
      \label{Eq:qm-rel-6}
\ea
These relations are valid in a large class of quark models, including 
spectator models, bag model, light-front constituent quark model 
\cite{Jakob:1997wg,Avakian:2009jt,Avakian:2010br,Pasquini:2008ax,Lorce:2011zta}.
The CPM also supports the following non-linear quark model relations
\ba
      h_1^q(x,p_T)\,h_{1T}^{\perp q}(x,p_T) &=& 
      -\,\frac12\biggl[h_{1L}^{\perp q}(x,p_T)\biggr]^2\label{Eq:qm-rel-7}\\
      g_T^q(x,p_T)\,g_{T}^{\perp q}(x,p_T) &=& 
      +\,\frac12\biggl[g_{1T}^{\perp q}(x,p_T)\biggr]^2
      -g_{1T}^{\perp q}(x,p_T)\,g_{L}^{\perp q}(x,p_T) \label{Eq:qm-rel-8}
\ea
\esub
In \cite{Efremov:2009ze} it was shown that the CPM complies with the 
relations (\ref{Eq:qm-rel-1},~\ref{Eq:qm-rel-4},~\ref{Eq:qm-rel-7}) 
featuring only twist-2 TMDs. Here we see that the model supports the
the full set of linear and non-linear quark model relations 
(\ref{Eq:quark-model-relations}).

One can impose the additional assumption of the SU(4) 
spin-flavor symmetry in the CPM by assuming 
${\cal G}^q(pP)=N^q{\cal F}(pP)$ and ${\cal G}^q(pP)=P^q{\cal F}(pP)$. 
For a proton the SU(4) spin flavor factors are given by \cite{Karl:1984cz}
\bsub
\label{Eq:Nq+Pq}
\ba\label{Eq:Nq} N_u = \frac{N_c+1}{2}\,, &&  N_d = \frac{\;N_c-1}{2}\,, \\
   \label{Eq:Pq} P_u = \frac{N_c+5}{6}\,, &&  P_d = \frac{-N_c+1}{6}\,,
\ea
\esub
and those for neutron follow from interchanging $u\leftrightarrow d$.
Under the assumption of SU(4) spin-flavor symmetry
and introducing the definition ${\cal D}^q=P^q/N^q$ also the following 
SU(4) quark model relations equations hold in the CPM
\bsub
\label{Eq:qm-rel-SU4}
\ba
      {\cal D}^q f_1^q(x,p_T)+g_1^q(x,p_T) &=& 2h_1^q(x,p_T), 
      \label{Eq:qm-rel-SU4a}\\
      {\cal D}^q e^q(x,p_T)+h_L^q(x,p_T) &=& 2g_T^q(x,p_T), 
      \label{Eq:qm-rel-SU4b}\\
      {\cal D}^q f^{\perp q}(x,p_T)       &=& h_T^{\perp q}(x,p_T).
      \label{Eq:qm-rel-SU4c}
\ea
\esub
It is important to stress that relations connecting unpolarized and 
polarized TMDs such as (\ref{Eq:qm-rel-SU4}) require the stronger
additional assumption of SU(4) spin-flavor symmetry.

The deeper reason for the appearance of the quark model relations 
including twist-2 TMDs,
(\ref{Eq:qm-rel-1},~\ref{Eq:qm-rel-4},~\ref{Eq:qm-rel-4}) 
(and (\ref{Eq:qm-rel-SU4a}) under the additional assumption of SU(4)
spin-flavor symmetry), can be traced back to the symmetries of the 
lightcone wave-functions in a large class of 
independent-particle models where the quarks do not interact 
with each other but are bound by a mean field \cite{Lorce:2011zta}.
Not all models support these relations. A counter-example are 
quark-target models \cite{Meissner:2007rx} where the relations
(\ref{Eq:quark-model-relations},~\ref{Eq:qm-rel-SU4}) are not valid.

\subsection{Wandzura-Wilczek relations}

In the CPM we can derive the following relations
from Eqs.~(\ref{Eq:EOM},~\ref{Eq:LIRs})
\bsub\label{Eq:WW}
\begin{align}
   g_T^q(x)  \; \stackrel{\rm WW}{=}& \phantom{2x}\; 
   \int_x^1\frac{\mathrm{d}y}{y}\,g_1^q(y) +\frac{m}{M}\,
   \biggl[-\,\frac{h_1^q(x)}{x}+\int_x^1\frac{\mathrm{d}y}{y^2}\,h_1^q(y)\,
   \biggr]\,,\label{Eq:WW1} \\
   h_L^q(x)  \; \stackrel{\rm WW}{=}& \;2x 
   \int_x^1\frac{\mathrm{d}y}{y^2}\,h_1^q(y) +\frac{m}{M}\,
   \biggl[\frac{g_1^q(x)}{x}-2x\int_x^1\frac{\mathrm{d}y}{y^3}\,g_1^q(y)
   \biggr]\,.\label{Eq:WW2} 
\end{align}
\esub
The Eq.~(\ref{Eq:WW1}) is obtained by integrating (\ref{Eq:EOM-gT}) over 
$p_T$ with $\tilde{g}_T^q(x)=0$, solving for $g_{1T}^{\perp(1)q}(x)$ and 
inserting the result into (\ref{eq:LIR1}) and finally integrating 
$\frac{\mathrm{d}}{\mathrm{d}x}\,g_T^q(x)$. Similarly Eq.~(\ref{Eq:WW2}) 
is derived from the EOM relation (\ref{Eq:EOM-hL}) and the LIR (\ref{eq:LIR2}).
As they follow from EOM relations and LIRs, the WW relations (\ref{Eq:WW}) 
contain no new information. 

From the point of view of QCD the relations (\ref{Eq:WW}) are of 
interest, as they constitute the so-called Wandzura-Wilczek approximation 
in QCD based on the neglect of tilde-terms (and in practice also mass terms) 
\cite{Wandzura:1977qf,Jaffe:1991ra}. This approximation is supported by 
instanton vacuum calculations \cite{Balla:1997hf,Dressler:1999hc}
where the tilde terms $\tilde{g}_T^q(x)$ and $\tilde{h}_L^q(x)$ are
small compared to the twist-2 terms in (\ref{Eq:WW}). 
The smallness of $\tilde{g}_T^q(x)$ is supported by
lattice QCD studies \cite{Gockeler:2000ja,Gockeler:2005vw}, and 
experiment \cite{Abe:1998wq,Anthony:2002hy,Airapetian:2011wu}
which indicate that $\tilde{g}_T^q(x)$ and $\tilde{h}_L^q(x)$ are
small compared to the twist-2 terms in (\ref{Eq:WW}). 
In the CPM the tilde terms are exactly zero, and
the relations (\ref{Eq:WW}) hold exactly.

For completeness we remark that the assumption that tilde-terms (and mass 
terms) are small and numerically negligible has been also applied to TMDs 
\cite{Efremov:2001cz,Efremov:2001ia,Kotzinian:2006dw,Avakian:2007mv,
Metz:2008ib,Accardi:2009au,Bastami:2018xqd,Bastami:2020asv}.
In this case one speaks of Wandzura-Wilczek-type approximations.
Neglecting current quark mass effects, two examples of such relations 
are \cite{Avakian:2007mv}
\bsub\label{Eq:WW-type}\begin{align}
   g_{1T}^{\perp(1)q}(x)  \; \stackrel{\rm WW-type}{=}& \hspace{7mm} x\; 
   \int_x^1\frac{\mathrm{d}y}{y}\,g_1^q(y) \,,\label{Eq:WW1-type} \\
   h_{1L}^{\perp(1)q}(x)  \; \stackrel{\rm WW-type}{=}& \;-2x^2 
   \int_x^1\frac{\mathrm{d}y}{y^2}\,h_1^q(y)\,.\label{Eq:WW2-type} 
\end{align}\esub
Such approximations can be tested in SIDIS and Drell-Yan 
experiments \cite{Bastami:2018xqd,Bastami:2020asv,Koike:2008du}.
In the CPM the relations (\ref{Eq:WW-type}) are exact.

\subsection{Independent amplitudes in the covariant parton model}

The previous sections have shown that the CPM
supports many relations. It is an interesting question how many
independent TMDs exist in this model. For that we evaluate the
amplitudes of the unintegrated correlator (\ref{eq:corr-fully-unint}).
As expected, the T-odd $A_4^q$, $A_5^q$, $A_{12}^q$ are zero. For the
T-even amplitudes we obtain 
\bsub
\label{Eq:Ai-model}
\ba
  A_1^q &=&  P^0 \,\Theta(p^0)\,\delta(p^2 - m^2)\,{\mathcal{G}^q(pP)}
  \,\frac{m}{M} \, , \label{Eq:A1-model}\\
  A_2^q &=& 0 \, , \phantom{\frac11}
  \label{Eq:A2-model}\\
  A_3^q &=&  P^0 \,\Theta(p^0)\,\delta(p^2 - m^2)\,{\mathcal{G}^q(pP)}
   \, ,\label{Eq:A3-model}\phantom{\frac11}\\
  A_6^q &=&  P^0 \,\Theta(p^0)\,\delta(p^2 - m^2)\,{\mathcal{H}^q(pP)}
  \biggl(-\,\frac{m}{M}\biggr) \, , \label{Eq:A6-model}\\
  A_7^q &=&  P^0 \,\Theta(p^0)\,\delta(p^2 - m^2)\,{\mathcal{H}^q(pP)}
  \,\frac{mM}{p \cdot P + m M} \, ,\label{Eq:A7-model}\\
  A_8^q &=&  P^0 \,\Theta(p^0)\,\delta(p^2 - m^2)\,{\mathcal{H}^q(pP)}
  \,\frac{M^2}{p \cdot P + m M} \label{Eq:A8-model}\\
  A_9^q &=& 0 \,
  \label{Eq:A9-model}\\
  A_{10}^q &=&  P^0 \,\Theta(p^0)\,\delta(p^2 - m^2)\,{\mathcal{H}^q(pP)}
  \,\biggl(\,-1\,\biggr)
  \,, \label{Eq:A10-model}\\
  A_{11}^q &=& P^0 \,\Theta(p^0)\,\delta(p^2 - m^2)\,{\mathcal{H}^q(pP)}
  \,\biggl( - \frac{M^2}{p \cdot P + m M} \biggr) \,. \label{Eq:A11-model}
\ea
\esub
Interestingly the T-even amplitudes $A_2^q$, $A_9^q$ also vanish, 
because our model does not generate these Lorentz-structures.
The remaining 7 non-zero T-even amplitudes are related to each other
by 5 relations. In the unpolarized case we have 2 non-zero amplitudes 
related by 1 relation,
\be
     A_1^q = \frac{m}{M}\,A_3^q\,. \label{Eq:unp-ampl-model-rel}
\ee
In the polarized case we have 5 non-zero amplitudes related by 4 relations,
\ba
&&    A_6^q = -\,\frac{m}{M}\,A_{10}^q\,,\nonumber\\
&&    A_7^q = +\,\frac{m}{M}\,A_8^q\,,\nonumber\\
&&    A_8^q = -\,A_{11}^q\,,\nonumber\\
&&    A_{10}^q = -\;\frac{pP + mM}{M^2}\, A_8^q \,.
      \label{Eq:pol-ampl-model-REL}
\ea
Notice that for an onshell particle the relation
$P\cdot p = \frac{\vec{p}_T^{\;2} + m^2}{2x} + \frac12\,xM^2$ holds.
Thus in the model (but not in general) the product $ P\cdot p$
does not depend on $p^-$ and can be pulled out of the integrals
over $p^-$ in (\ref{Eq:TMDs-as-amp}).
Inserting the results (\ref{Eq:Ai-model}) for the amplitudes into
the expressions (\ref{Eq:TMDs-as-amp}) we recover the results obtained 
in Eqs.~(\ref{Eq:f1}, \ref{Eq:g1-g1Tperp}, \ref{Eq:TMD-twist2-chiral-odd}, 
\ref{eq:result-twist-3})
which provides an independent test of the model.

\subsection{Independent TMDs in the model}

The previous section has shown that there are two independent amplitudes: 
one unpolarized and one polarized. Consequently, all unpolarized TMDs are 
related to each other, and all polarized TMDs are related to each other.
Thus, it is possible to choose one unpolarized and one polarized TMD,
and express all other TMDs in terms of them.

In the unpolarized case we have 3 T-even TMDs $f_1^q(x,p_T)$, 
$e^q(x,p_T)$, $f^{\perp q}(x,p_T)$. We can choose the well-known 
unpolarized twist-2 TMDs as basis function. The other unpolarized
T-even TMDs are then given by
\bsub
\label{Eq:model-basis-f1}
\ba
         \label{Eq:model-e-rel}
         xe^q(x,p_T) \; &\stackrel{\rm CPM}{=}& \frac{m}{M}\,f_1^q(x,p_T) \\ 
         \label{Eq:model-fperp-rel}
         xf^{\perp q}(x,p_T) \; &\stackrel{\rm CPM}{=}& f_1^q(x,p_T) 
\ea
\esub
Notice that $e^q(x,p_T)$ vanishes if we one neglects current quark
mass effects.

In the polarized T-even sector we have 11 TMDs. At first glance it would seem
natural to express all polarized TMDs in terms of the relatively well-known 
helicity distribution $g_1^q(x,p_T)$. This is possible, but not ideal for 
the following reason. In the CPM, $g_1^q(x,p_T)$ exhibits 
a node at $p_T=x\,M$ (neglecting current quark masses) \cite{Efremov:2010mt}. 
Consequently TMDs without nodes would be expressed in terms $g_1^q(x,p_T)$ 
divided by a prefactor which is singular at $p_T=x\,M$ in order to remove 
the node present in $g_1^q(x,p_T)$. This is impractical for phenomenological 
applications. 

It is clearly an advantage to choose a TMD without a node to express the
other TMDs. A convenient choice for a basis function for polarized TMDs
is transversity $h_1^q(x,p_T)$ which exhibits no node in the model and is, 
after $g_1^q(x,p_T)$, the currently best known T-even TMD.
It is convenient to neglect current quark mass effects which make the
explicit expressions quite bulky and can be safely expected to be small 
in phenomenological applications.

%
%

It is convenient to quote the results in 3 groups in which the 
polarized TMDs have structurally similar expressions. We can express
$g_1^q(x,p_T)$, $g_L^{\perp q}(x,p_T)$, $h_T^{\perp q}(x,p_T)$ (which exhibit 
nodes) and $h_T^{\perp q}(x,p_T)$ in terms of transversity as
\bsub
\label{Eq:model-basis-h1}
\begin{align}
   g_1^q(x,p_T) \,  \stackrel{\rm CPM}{=} \, \biggl[1-\frac{p_T^2}{x^2M^2}\biggr]
   \;h_1^q(x,p_T) & \, , \label{Eq:model-g1-rel}\\
   g_L^{\perp q}(x,p_T) \,  \stackrel{\rm CPM}{=}  \, \frac1x\, \biggl[1-\frac{p_T^2}{x^2M^2}\biggr]
   \;h_1^q(x,p_T) & \, , \label{Eq:model-gLperp-rel}\\
   h_T^q(x,p_T) \,  \stackrel{\rm CPM}{=}  \,-\, \frac1x\,\biggl[1-\frac{p_T^2}{x^2M^2}\biggr]
   \;h_1^q(x,p_T) & \, , \label{Eq:model-hT-rel}\\
   h_T^{\perp q}(x,p_T) \,  \stackrel{\rm CPM}{=}  \, \frac1x\, \biggl[1+\frac{p_T^2}{x^2M^2}\biggr]
   \;h_1^q(x,p_T) & \, . \label{Eq:model-hTperp-rel}
\end{align}
The TMDs $g_T^q(x,p_T)$ and $h_L^q(x,p_T)$ are expressed 
in terms of transversity (for $m=0$) as follows
\begin{align}
   g_T^q(x,p_T) \,  \stackrel{\rm CPM}{=}  \, \biggl[\frac{p_T^2}{x^2M^2}\biggr] h_1^q(x,p_T) & \, ,\\
   h_L^q(x,p_T) \,  \stackrel{\rm CPM}{=}  \, \biggl[\frac{2p_T^2}{x^2M^2}\biggr] h_1^q(x,p_T) & \, .
\end{align}
Notice the relation $h_L^q(x,p_T)  \stackrel{\rm CPM}{=} 2g_T^q(x,p_T)$ which holds for 
$m=0$. Finally, the Mulders-Kotzinian TMDs
$g_{1T}^{\perp q}(x,p_T)$ and $h_{1L}^{\perp q}(x,p_T)$, pretzelosity
$h_{1T}^{\perp q}(x,p_T)$, and twist-3 TMD $g_T^{\perp q}(x,p_T)$
are expressed in terms of transversity as
\begin{align}
   g_{1T}^{\perp q}(x,p_T) \,  \stackrel{\rm CPM}{=}  \, \frac2x\,h_1^q(x,p_T)& \, ,\\
   h_{1L}^{\perp q}(x,p_T) \,  \stackrel{\rm CPM}{=}  \, -\,\frac2x\,h_1^q(x,p_T)& \, , \\
   h_{1T}^{\perp q}(x,p_T) \,  \stackrel{\rm CPM}{=}  \, -\,\frac2{x^2}\,h_1^q(x,p_T) & \, ,\\
   g_T^{\perp q}(x,p_T) \,    \stackrel{\rm CPM}{=}  \,\frac2{x^2}\,h_1^q(x,p_T) & \, . 
\end{align}
\esub
It will be interesting to test these CPM predictions in future,
when more about TMDs will be known. 
These relations can also be investigated in other models.
It would be interesting to assess in this way whether the relations
(\ref{Eq:model-basis-f1}--\ref{Eq:model-basis-h1}) are supported by
other quark models and if so, within which accuracy.

\section{Numerical results}
\label{Sec-6:results}

In this section we show the numerical results. After a brief
review how the covariant functions ${\cal G}^a(pP)$ and ${\cal H}^a(pP)$ 
are obtained from the input PDFs $f_1^a(x)$ and $g_1^a(x)$,
we present predictions for unpolarized and polarized PDFs or transverse
moments of TMDs, and compare to parametrizations where available.

\subsection{Covariant functions and input PDFs}

The covariant functions ${\cal G}^a(pP)$
for $a=u,\,d,\,\bar{u},\,\bar{d},\,\dots$ are uniquely determined 
from respectively $f_1^a(x)$ and $g_1^a(x)$. With the notation
${\cal G}^q(p^0)$ and ${\cal H}^q(p^0)$ in the nucleon rest frame,
the relations are given by  \cite{Efremov:2010mt}
\ba\label{Eq:invert-PDFs}
     {\cal G}^q(p^0)\biggl|_{p^0=\frac12xM} &=& -\frac{1}{\pi M^3}
     \frac{d\;}{dx}\,\biggl[\frac{f_1^q(x)}{x}\biggr] , \nonumber\\
     {\cal H}^q(p^0)\biggl|_{p^0=\frac12xM} &=& \frac{1}{\pi M^3x^2}\Biggl[
     2\int_x^1\frac{dy}{y}\,g_1^q(y)+3g_1^q(x)-
     x\frac{d\;}{dx}\,\biggl[\frac{g_1^q(x)}{x}\biggr]\Biggr] .
\ea
Several comments are in order. First, the model relates different TMDs 
obeying different evolution equations to the same covariant functions.
Therefore, a scale must be chosen at which the covariant functions in 
(\ref{Eq:invert-PDFs}) are determined. The choice of this scale is 
part of the modeling. The renormalization scale, which is not indicated 
in (\ref{Eq:invert-PDFs}) for brevity, must be chosen large enough 
for the partonic picture to be justified, but is otherwise not fixed. 
In this work we choose $\mu^2=2.5\,{\rm GeV}^2$ which is a convenient 
scale because many extractions of TMDs from SIDIS data have been 
performed at comparable scales. 
Second, in the case of TMDs we strictly speaking deal with a 
double-scale problem, and the choice of both scales is part of the model.
The second scale, associated with the removal of rapidity divergences,
can be also chosen to be $\mu^2=2.5\,{\rm GeV}^2$. The dependence on 
this second scale is governed by the CSS evolution equations. 
Third, an important feature of the parton model is the partonic interpretation.
In QCD the partonic interpretation is strictly speaking only justified at 
leading-order (LO) (and for PDFs of the nucleon, but e.g.\ not for
nuclei \cite{Brodsky:2002ue}). For our calculations we therefore choose 
LO parametrizations. In order to investigate the dependence on the 
chosen input parametrization, we use several parametrizations for
respectively $f_1^a(x)$ and $g_1^a(x)$.

\begin{figure}[b!]
  \centering
  \includegraphics[height=4cm]{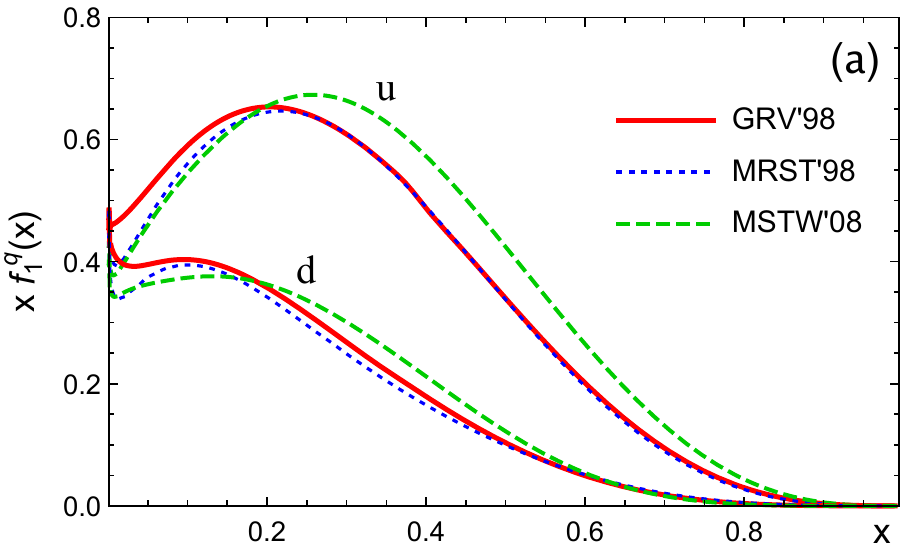} \hspace{1cm}
  \includegraphics[height=4cm]{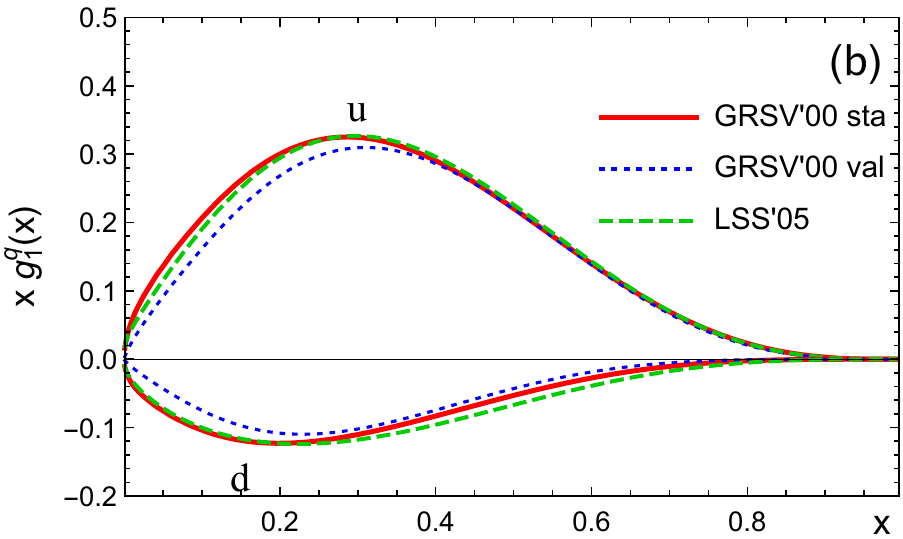} 
  \caption{\label{Fig01:f1-g1} 
The input PDFs used in this work for the model calculations 
as functions of $x$ at the scale $2.5$ GeV$^2$.
(a) LO parametrizations of the unpolarized PDFs $xf_1^q(x)$ from 
GRV'98 \cite{Gluck:1998xa}, 
MRST'98 \cite{Martin:1998sq}
MSTW'09 \cite{Martin:2009iq}.
(b) LO parametrizations of the helicity PDFs $xg_1^q(x)$ from
GRSV'00 standard and valence scenario (see text) \cite{Gluck:2000dy},
and LSS'05 \cite{Leader:2005ci}. }
\end{figure}

The parametrizations used in this work are shown in Fig.~\ref{Fig01:f1-g1}:
GRV'98 \cite{Gluck:1998xa}, MRST'98 \cite{Martin:1998sq}, 
MSTW'09 \cite{Martin:2009iq} for $f_1^q(x)$; standard and valence 
scenarios of GRSV'00 \cite{Gluck:2000dy}, LSS'05 \cite{Leader:2005ci} 
for $g_1^q(x)$. More recent parametrizations are available, e.g.\ 
\cite{Harland-Lang:2014zoa,Alekhin:2017kpj,Ball:2017nwa,Sato:2019yez,Hou:2019efy}
for unpolarized PDFs, or 
\cite{deFlorian:2014yva,Nocera:2014gqa,Leader:2014uua,Ethier:2017zbq,deFlorian:2019zkl}
for helicity PDFs, see Ref.~\cite{Ethier:2020way} for a review.
The reason why for our purposes the earlier parametrizations
\cite{Gluck:1998xa,Martin:1998sq,Martin:2009iq,Gluck:2000dy,Leader:2005ci} 
are preferable is because all more recent helicity PDF parametrizations 
were performed at next-to-leading order, and \cite{Gluck:2000dy,Leader:2005ci} 
are among the last LO helicity parametrizations (recall that the use of
LO parametrization is preferable in our partonic approach). 
The GRSV'00 \cite{Gluck:2000dy} parametrizations of $g_1^q(x)$ were
obtained using GRV'98 \cite{Gluck:1998xa} for $f_1^q(x)$, while 
LSS'05 parametrizations of $g_1^q(x)$ were obtained using $f_1^q(x)$ 
from MRST'98 \cite{Martin:1998sq}. In addition, we use also the more 
recent MSTW'09 \cite{Martin:2009iq} parametrizations for $f_1^q(x)$.

Over the last 2 decades the parametrizations of quark PDFs $f_1^q(x)$ 
and $g_1^q(x)$ for $q=u,\,d$ have changed moderately, unlike especially 
antiquark helicity parametrizations which changed significantly due to 
recent data and may change further due to future Drell-Yan data from BNL 
or the Electron-Ion Collider \cite{Aschenauer:2020pdk}.

\subsection{The \boldmath $p_T$-dependence of TMDs}

The $p_T$-dependence of twist-2 TMDs was discussed in 
\cite{Efremov:2010mt}.\footnote{At this occasion 
  we would like to make a correction regarding \cite{Efremov:2010mt}. 
  The analytical results in \cite{Efremov:2010mt} are correct 
  (except for obvious misprints in Eq.~(18) where it should be 
  $\xi\frac{dg_1^q(\xi)}{d\xi}$ on the right-hand-side instead of 
  $x\frac{dg_1^q(\xi)}{d\xi}$ and Eq.~(19) of \cite{Efremov:2010mt} 
  where it should be $M^2$ in the denominator instead of $M^3$).
  Also the numerical results  in \cite{Efremov:2010mt} were correctly 
  computed. But due to an unfortunate plotting mistake the Figs.~1-4 of 
  \cite{Efremov:2010mt} show exactly {\it half} of the correct results 
  for all the TMDs. None of the conclusions of \cite{Efremov:2010mt} 
  is affected by this mistake.}  
The new results for twist-3 TMDs derived here have very similar 
$p_T$-dependencies, and we can refrain from discussing them in this work. 
Instead, we will content ourselves with briefly reviewing the main features 
of the $p_T$-dependencies of TMDs in this section.

It is remarkable that the $p_T$-dependencies of $f_1^q(x,p_T)$ and 
other unpolarized T-even TMDs are uniquely predicted from the input 
PDF $f_1^q(x)$. Similarly, the $p_T$-dependencies of $g_1^q(x,p_T)$ 
and other polarized T-even TMDs are uniquely predicted from the input 
PDF $g_1^q(x)$.
This is possible due to the strong model assumption of onshell
quarks which leads to the 3D rotational symmetry in nucleon rest frame 
encoded in the covariant functions ${\cal G}^q(p^0)$, ${\cal H}^q(p^0)$ 
with $p^0=|\vec{p}|$ for massless quarks. This 3D symmetry connects 
longitudinal and transverse parton motion. 

The TMDs have finite support for $0<p_T<M\sqrt{x(1-x)}$ 
and vanish outside this range \cite{Zavada:1996kp}.
The covariant functions are non-zero only for $0 < p^0 < \frac12M$ 
for massless partons due to $0<x< 1$ in Eq.~(\ref{Eq:invert-PDFs}).
In this work we will restrict ourselves to the discussion
of model predictions after the transverse momenta are integrated out,
and present results for PDFs like $h_1^q(x)$ or $g_T^q(x)$ or 
transverse moments of TMDs, defined e.g.\ as
\be
       h_{1T}^{\perp (n)q}(x) = \int d^2p_T\,\biggl(\frac{p_T^2}{2M^2}\biggr)^{\!n}
       h_{1T}^{\perp q}(x,p_T)\,
\ee
and analog for other TMDs. It is convenient to describe the respective
structure functions processes in terms of such transverse moments. For 
instance in SIDIS the functions $h_1^q(x)$, $h_{1L}^{\perp (1)q}(x)$, 
$h_{1T}^{\perp (2)q}(x)$ enter \cite{Bastami:2018xqd}. In the following 
we show, unless otherwise stated, those functions (PDFs or certain 
transverse moments) which are relevant for phenomenological applications. 
Where possible we will test the model predictions by comparing to 
parametrizations.

\begin{figure}[h!]
  \centering
  \includegraphics[height=4cm]{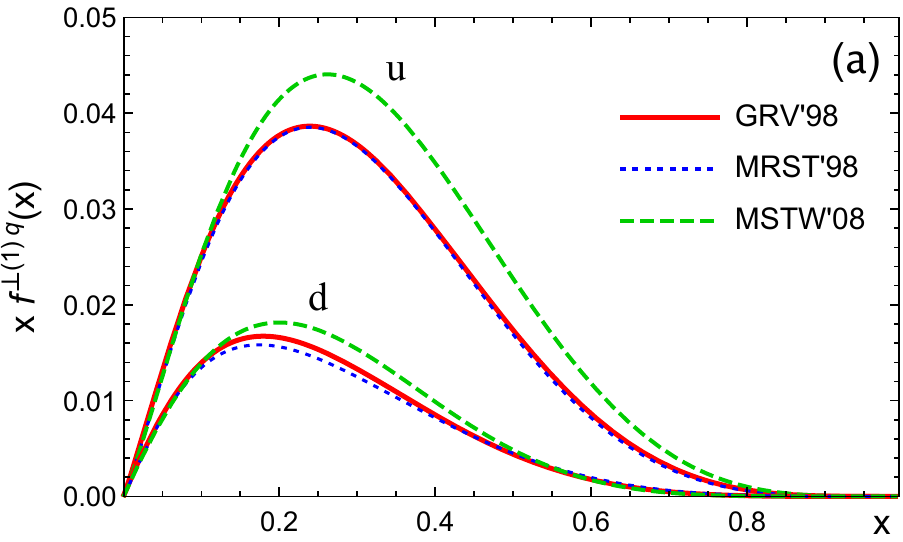} \hspace{1cm}
  \includegraphics[height=4cm]{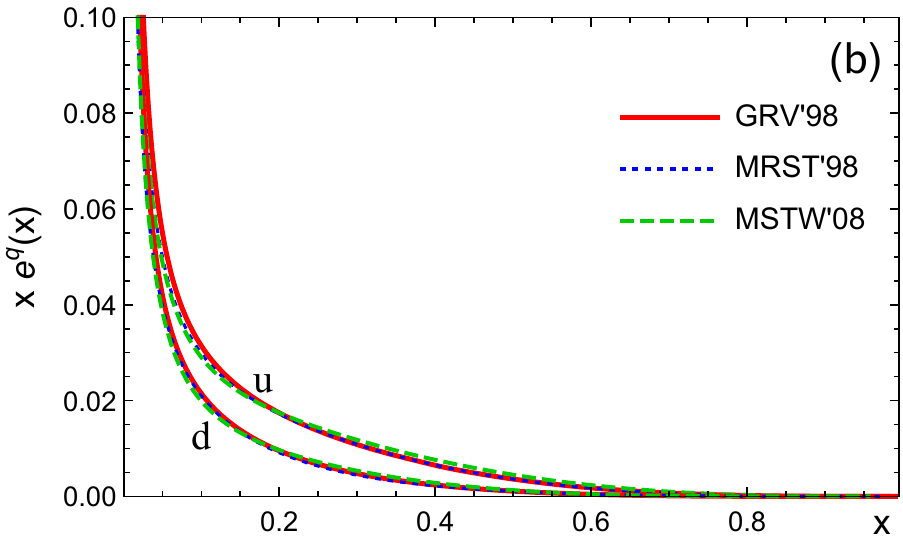} 
  \caption{\label{Fig02:fperp-e} 
Predictions for (a) $xf^{\perp(1)q}(x)$ and (b) $xe^q(x)$ as functions 
of $x$ at the scale $2.5$ GeV$^2$ from the CPM according to 
Eqs.~(\ref{Eq:model-e-rel},~\ref{Eq:model-fperp-rel}) using the 
unpolarized LO input PDFs shown in Fig.~\ref{Fig01:f1-g1}a.}
\end{figure}

\subsection{Predictions for unpolarized TMDs}

In the unpolarized sector there are only 3 T-even TMDs:
the twist-2 $f_1^q(x,p_T)$ which is input in the model, and 
the twist-3 $f^{\perp q}(x,p_T)$ and $e^q(x,p_T)$. Both TMDs
are related in the CPM to $f_1^q(x,p_T)$ according to 
Eqs.~(\ref{Eq:model-e-rel},~\ref{Eq:model-fperp-rel}).

No relation between $f_1^q(x,p_T)$ and $f^{\perp q}(x,p_T)$ exists 
in QCD due to the appearance of the function $\tilde{f}^\perp(x,p_T)$.
In general in quark models the tilde terms are also non-zero due to 
non-trivial quark-model interactions. It is therefore interesting that 
the relation (\ref{Eq:model-fperp-rel}) holds in lightfront constituent 
quark model and chiral quark soliton model even though both models exhibit 
nontrivial model interactions, encoded in the nonperturbative lightfront 
wave functions of the former or provided by the strong chiral interactions 
of the latter model \cite{Lorce:2014hxa}. 

The CPM prediction for $xf^{\perp(1)q}(x)$ is shown in Fig.~\ref{Fig02:fperp-e}a,
and corresponds exactly to the estimate for this TMD obtained in 
\cite{Bastami:2018xqd} on the basis of the WW-type approximation. 
The result in Fig.~\ref{Fig02:fperp-e}a shows that 
$xf^{\perp(1)q}(x)$ is sizable. 

The situation is different for $xe^q(x)$ which vanishes if we neglect
quark mass effects. In order to show a non-zero result we assume 
$m_q=5\,{\rm MeV}$ for both $u$- and $d$-flavor. As expected, the 
$xe^q(x)$ resulting from Eq.~(\ref{Eq:model-e-rel}) is very small. 
Assuming TMD factorization at twist-3 level \cite{Gamberg:2006ru}, 
$e^q(x,p_T)$ contributes to observables e.g.\ in SIDIS with the prefactor 
$M/Q$ \cite{Bacchetta:2006tn} where $Q$ is the hard scale of the process. 
The contribution of the mass term in $e^q(x,p_T)$ is therefore effectively
proportional to $m_q/Q$ and can be safely 
neglected in many phenomenological applications \cite{Bastami:2018xqd}. 
Only at extremely small $x \lesssim m_q/Q$ could the mass term contribution 
to $xe^q(x)$ become important. This kinematics will be accessible at the
Electron-Ion Collider \cite{Accardi:2012qut}.
However, the discussion of TMDs at small $x$ is more adequately addressed
in the approach of Refs.~\cite{Kovchegov:2018znm,Kovchegov:2018zeq}
which, to the best of our knowledge, has not yet been applied
to subleading twist. It is interesting to remark that e.g.\ in the
lightfront constituent quark model the relation (\ref{Eq:model-e-rel}) 
is also valid, but $e^q(x)$ in nevertheless sizable, because that model
operates at a low hadronic scale $\mu < 1\,{\rm GeV}$ where
the effective quark degrees of freedom have a constituent quark
mass of about $300\,{\rm MeV}$ \cite{Lorce:2014hxa}.

No model-independent extractions of these twist-3 TMDs are currently 
available. In the case of $e^q(x)$ very first (and model-dependent)
extractions were reported in Refs.~\cite{Efremov:2002ut,Courtoy:2014ixa}.
For further model studies of $f^{\perp(1)q}(x)$ and $e^q(x)$, including 
the interesting possibility of a singular $\delta(x)$-contribution 
to $e^q(x)$ which is beyond our partonic approach, we refer to 
Refs.~\cite{Jaffe:1991ra,Jakob:1997wg,Burkardt:2001iy,Efremov:2002qh,
Schweitzer:2003uy,Wakamatsu:2003uu,Ohnishi:2003mf,Cebulla:2007ej,
Avakian:2009jt,Avakian:2010br,Mukherjee:2009uy,Mukherjee:2010iw,
Pasquini:2018oyz,Aslan:2018tff,Ma:2020kjz,Bhattacharya:2020jfj}. 
Noteworthy is the partonic interpretation of the pure twist-3 
contribution to $e^q(x)$ in terms of transverse forces experienced
by quarks in DIS \cite{Burkardt:2008ps}.

\begin{figure}[b!]
  \centering
  \includegraphics[height=4cm]{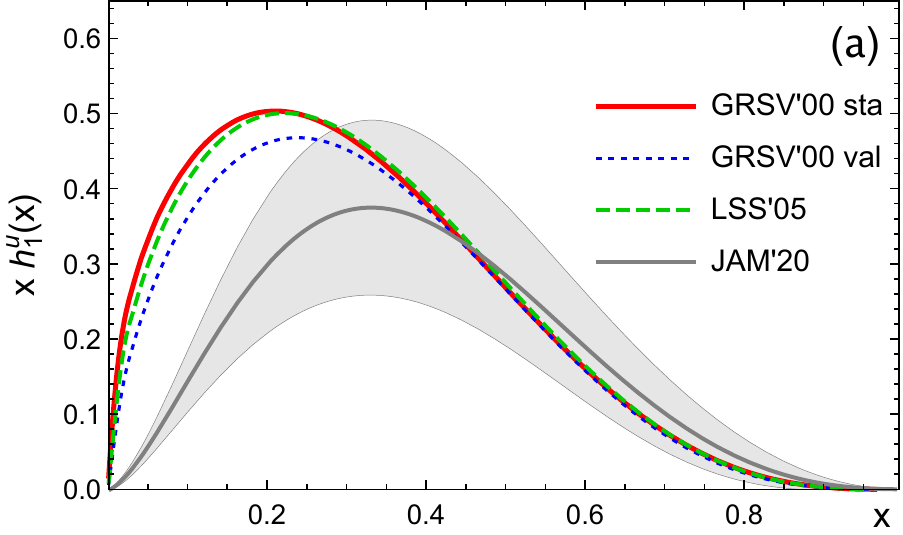} \hspace{1cm}
  \includegraphics[height=4cm]{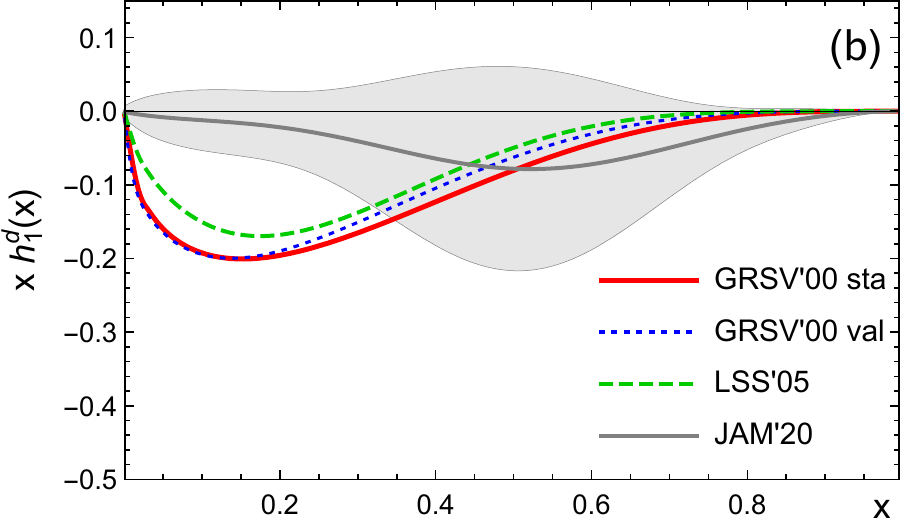}

\vspace{1cm}

  \includegraphics[height=4cm]{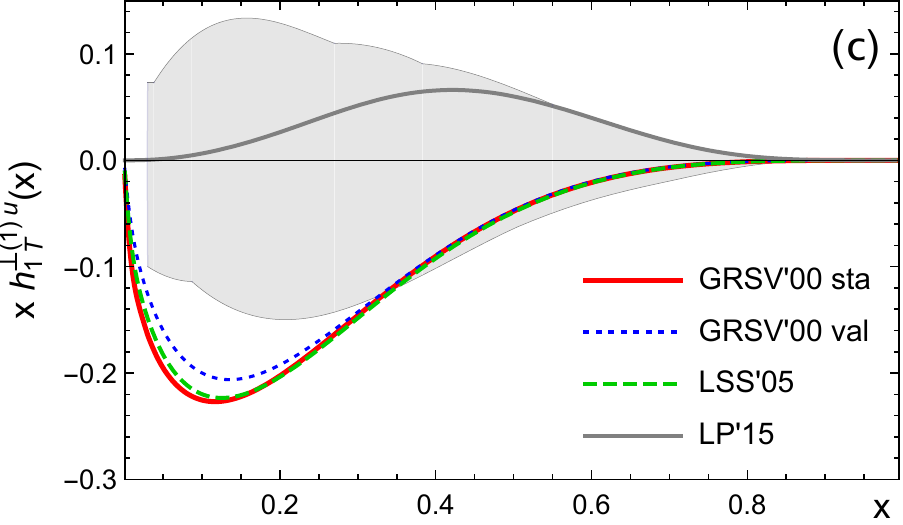} 
  \hspace{1cm}
  \includegraphics[height=4cm]{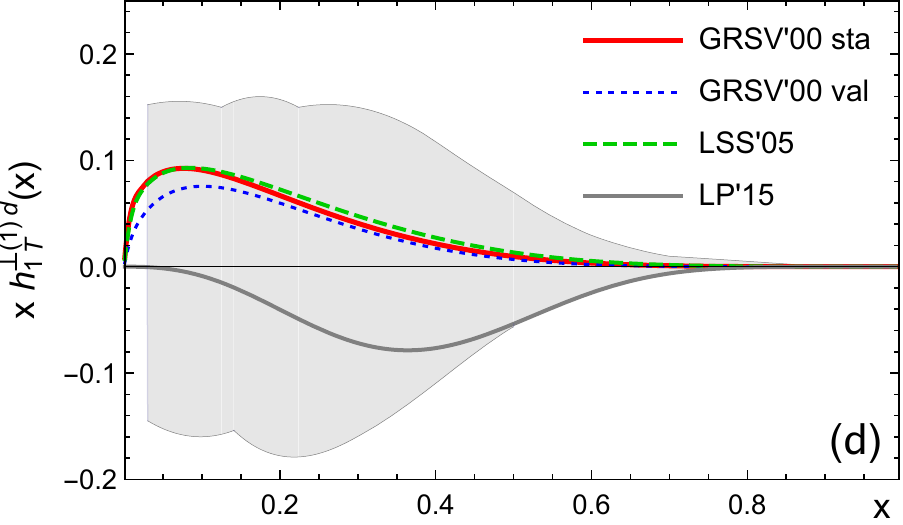} 

  \caption{\label{Fig03:h1-h1Tperp1}
Model predictions for $xh_1^q(x)$, (a) and (b), and $xh_{1T}^{\perp(1)q}(x)$, 
(c) and (d), as functions of $x$ at the scale $2.5$ GeV$^2$ obtained using 
the helicity LO input PDFs from Fig.~\ref{Fig01:f1-g1}b in comparison to 
respectively the JAM'20 \cite{Cammarota:2020qcw} and LP'15 \cite{Lefky:2014eia} 
parametrizations.}
\end{figure}

\subsection{Results for transversity and pretzelosity}

Next we turn our attention to polarized TMDs starting the discussion 
with the twist-2 transversity and pretzelosity which can be compared 
to available parametrizations \cite{Cammarota:2020qcw,Lefky:2014eia}.

In Figs.~\ref{Fig03:h1-h1Tperp1}a and ~\ref{Fig03:h1-h1Tperp1}b we compare 
the model predictions for $xh_1^q(x)$ to the recent JAM'20 parametrization 
\cite{Cammarota:2020qcw}. The CPM describes the sign and magnitude of the 
transversity quark distributions well. For larger $x\gtrsim0.2$ the 
quantitative agreement is very good and the model results are close to 
or within the 1-$\sigma$ region of the extraction \cite{Cammarota:2020qcw}.
At smaller $x\lesssim 0.2$ the model has a tendency to overestimate the
JAM'20 parametrization for $u$ and $d$ flavors. The uncertainty of
the JAM'20 parametrization \cite{Cammarota:2020qcw} is still very large.
For instance the $d$-quark transversity is compatible with zero within the 
1-$\sigma$ uncertainty of the extraction. Future data will constrain more 
strongly the extractions and allow us to test the model predictions for 
$h_1^q(x)$ more rigorously. The CPM is in good qualitative agreement with 
other model calculations 
\cite{Jaffe:1991ra,Jakob:1997wg,Avakian:2009jt,Avakian:2010br,Pasquini:2008ax,
Cloet:2007em,Bacchetta:2008af,Maji:2017bcz,Gamberg:1998vg,Wakamatsu:2000fd,
Schweitzer:2001sr,Yazdi:2014zaa,Maji:2016yqo,Kofler:2017uzq,
Xu:2019xhk,Barone:2001sp} and lattice QCD 
\cite{Chen:2016utp,Alexandrou:2016jqi}.

In Figs.~\ref{Fig03:h1-h1Tperp1}c and ~\ref{Fig03:h1-h1Tperp1}d we 
compare the model predictions for $xh_{1T}^{\perp(1)q}(x)$ to the LP'15
fit \cite{Lefky:2014eia} 
(where the (1)-moment was extracted, though in phenomenological applications 
\cite{Bastami:2018xqd,Bastami:2020asv} the (2)-moment of pretzelosity enters
naturally).
The present data on the azimuthal asymmetry related to pretzelosity are
compatible with zero, which is reflected by the uncertainty band of 
the LP'15 parametrization for $u$ and $d$ flavors \cite{Lefky:2014eia}. 
The best fit of LP'15 has opposite sign to the CPM which should be not
too disturbing considering the large uncertainties of the fit. The CPM 
results agree with other models
\cite{Jakob:1997wg,Avakian:2009jt,Avakian:2010br,Pasquini:2008ax,Bacchetta:2008af,She:2009jq,Lu:2012gu,Maji:2015vsa,Maji:2016yqo,Maji:2017bcz}.
More precise future data are needed to test the model predictions for 
pretzelosity.
This TMD is of interest because it is related to deviations of the nucleon's
transverse spin distribution from spherical symmetry \cite{Miller:2007ae} and,
in certain models, to quark orbital momentum 
\cite{Avakian:2008dz,She:2009jq,Lorce:2011kn}.

\subsection{Kotzinian-Mulders functions}

In this Section we continue the discussion of polarized twist-2 TMDs 
for which currently no extractions are available, the Kotzinian-Mulders 
functions $g_{1T}^{\perp(1)q}(x)$ and $h_{1L}^{\perp(1)q}(x)$. 
In the CPM and in many other models these TMDs are related to each 
other by the Eq.~(\ref{Eq:qm-rel-1}). For clarity we nevertheless 
show the results for both TMDs in separate figures.

\begin{figure}[h!]
  \centering
  \vspace{5mm}
  \includegraphics[height=4cm]{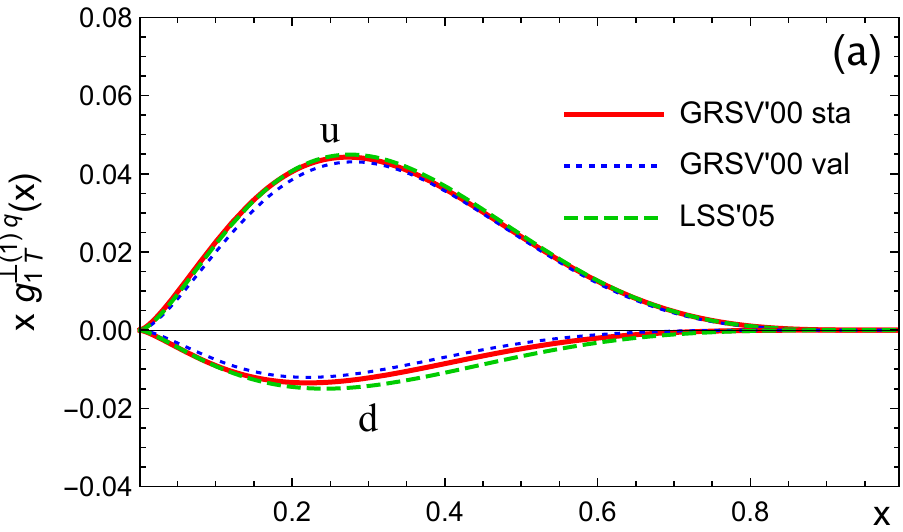} \hspace{1cm}
  \includegraphics[height=4cm]{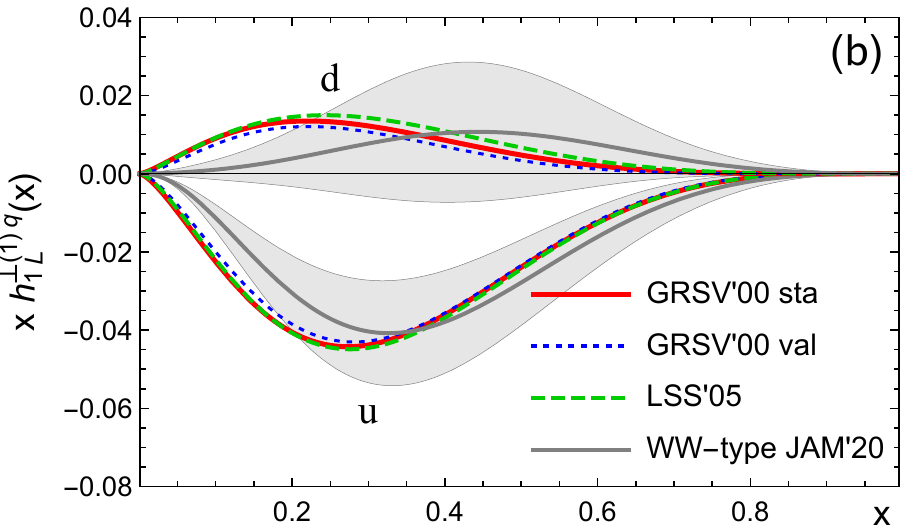}
  \caption{\label{Fig04:gear-worms} 
Model predictions for (a) $xg_{1T}^{\perp(1)q}(x)$ and (b) $xh_{1L}^{\perp(1)q}(x)$ 
for $q=u,\,d$ as functions of $x$ at the scale $2.5$ GeV$^2$ obtained using 
the helicity LO input PDFs from Fig.~\ref{Fig01:f1-g1}b. For comparison we
show in (b) the estimate for $xh_{1L}^{\perp(1)q}(x)$ from \cite{Bastami:2018xqd} 
based on the WW-type approximation and JAM'20 transversity parametrization 
\cite{Cammarota:2020qcw}.}
\end{figure}

In Fig.~\ref{Fig04:gear-worms}a we show the model results for 
$xg_{1T}^{\perp(1)q}(x)$. The model supports the WW-type relation 
(\ref{Eq:WW1-type}) which was used in Ref.~\cite{Bastami:2018xqd}.
This means that in Ref.~\cite{Bastami:2018xqd} exactly the same predictions 
as presented in Fig.~\ref{Fig04:gear-worms}a where used for this TMD, and 
shown to be compatible with the data currently available on this TMD. 
This means that the CPM model prediction for $g_{1T}^{\perp(1)q}(x)$ is
also compatible with the currently available SIDIS data. One should
add that the existing data on the pertinent SIDIS asymmetry have 
sizable error bars and this test of the model is at the current stage
rather qualitative. However, more precise future data will allow us to
make more quantitative tests of the model.

In Fig.~\ref{Fig04:gear-worms}b we plot the model results for 
$xh_{1L}^{\perp(1)q}(x)$. For comparison we show also the estimate for 
this TMD from \cite{Bastami:2018xqd} which is based on the WW-type 
approximation (\ref{Eq:WW2-type}) and JAM'20 $h_1^q(x)$ parametrization 
\cite{Cammarota:2020qcw}. Though the CPM supports Eq.~(\ref{Eq:WW2-type}), 
the comparison in Fig.~\ref{Fig04:gear-worms}b is nevertheless interesting: 
in our model this TMD is ultimately obtained from the input helicity PDF. 
In contrast to this the WW-type-based prediction for $h_{1L}^{\perp(1)q}(x)$ 
from \cite{Bastami:2018xqd} is based on (\ref{Eq:WW2-type}) and transversity 
as input. Thus, these are two different ways of making predictions for
this Kotzinian-Mulders function and the good agreement of the two results 
constitutes a consistency check in the sense that the CPM practically 
supports numerical estimates of this TMD based on the WW-type 
approximations \cite{Bastami:2018xqd}.

\subsection{Predictions for polarized twist-3 PDFs and TMDs}

\begin{figure}[t!]
  \centering
  \includegraphics[height=4cm]{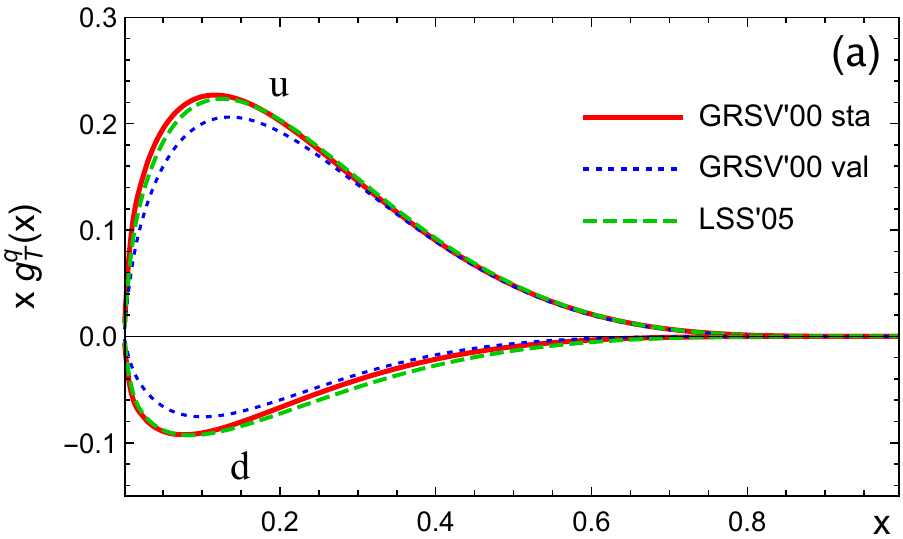} \hspace{1cm}
  \includegraphics[height=4cm]{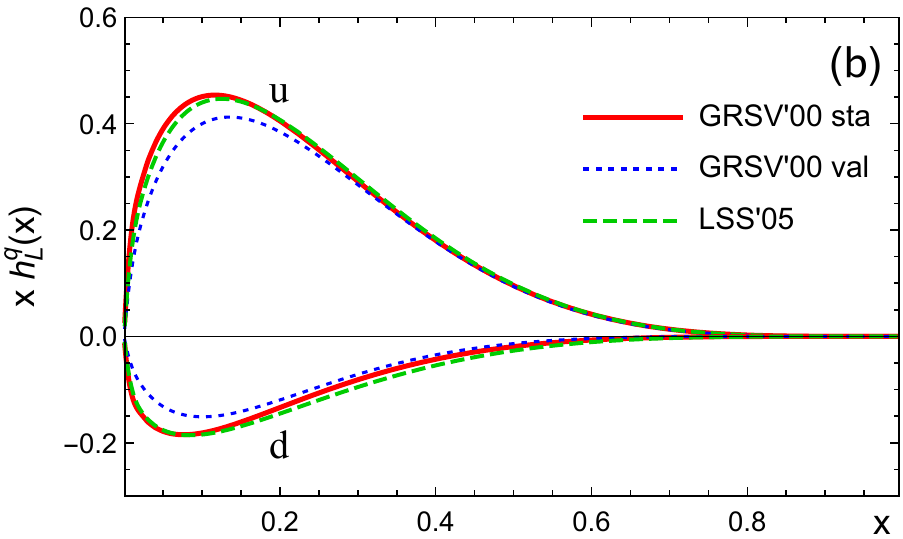}
  \caption{\label{Fig05:gT-hL} 
Model predictions for the twist-3 PDFs (a) $xg_T^q(x)$ and (b) $xh_L^q(x)$ 
as functions of $x$ at the scale $2.5$ GeV$^2$ obtained using the 
helicity LO input PDFs from Fig.~\ref{Fig01:f1-g1}b.}
\end{figure}

In this section we discuss polarized twist-3 TMDs beginning with 
$g_T^q(x)$ which is accessible in polarized DIS making it the only 
well-constrained twist-3 function. In the CPM the WW-approximation 
(\ref{Eq:WW1}) holds exactly. The model prediction is shown in 
Fig.~\ref{Fig05:gT-hL}a.
Interestingly, the Mellin moments of the pure twist-3 contribution
$\tilde{g}_T^q(x)$ were shown in instanton vacuum calculations to be 
strongly suppressed by powers of the instanton packing fraction
\cite{Balla:1997hf}. Subsequently the smallness of the $\tilde{g}_T^q(x)$ 
contribution to $g_T^q(x)$ was confirmed in experiments 
\cite{Abe:1998wq,Anthony:2002hy,Airapetian:2011wu} and 
lattice QCD studies \cite{Gockeler:2000ja,Gockeler:2005vw}.
These theoretical and experimental results have been a main motivation 
for exploring the possibility of WW- and WW-type approximations
\cite{Metz:2008ib,Anselmino:2013vqa,Bastami:2018xqd,Bastami:2020asv}
though it cannot be excluded that $\tilde{g}_T^q(x)$ might be sizable 
in certain (so far experimentally unexplored or poorly unconstrained) 
$x$-regions \cite{Accardi:2009au}.
The smallness of $\tilde{g}_T^q(x)$ provides important support also for
the CPM, and played an important role in the development of 
this model \cite{Zavada:2002uz}.

The only further polarized twist-3 collinear PDF is $h_L^q(x)$
which, however, is chirally odd and not accessible in DIS. 
Consequently almost nothing is known phenomenologically about
this PDF. Interestingly, also in this case the pure twist-3 
contribution $\tilde{h}_L^q(x)$ is also suppressed in the 
instanton vacuum \cite{Dressler:1999hc}. In the CPM the corresponding 
WW-relation (\ref{Eq:WW2}) is exact. The model prediction for $xh_L^q(x)$ 
is shown in Fig.~\ref{Fig05:gT-hL}b. 
This function contributes to single spin asymmetries in SIDIS 
\cite{Mulders:1995dh} with several other unknown twist-3 TMDs and 
fragmentation functions such that phenomenological information on this
TMD is difficult to obtain \cite{Efremov:2001cz,Efremov:2001ia}. Of 
interest is the CPM prediction $h_L^q(x)=2g_T^q(x)$. It will be 
interesting to see if this prediction will be supported phenomenologically.

\begin{figure}[t!]
  \includegraphics[height=4cm]{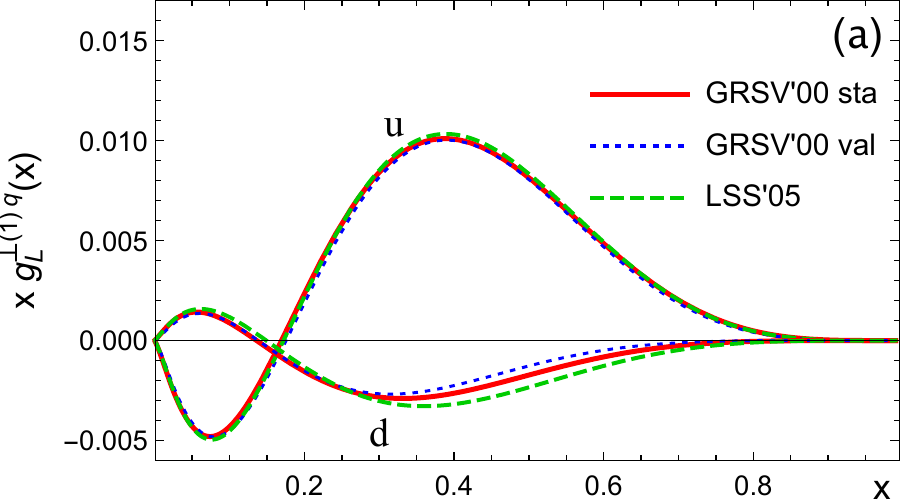} \hspace{1cm}
  \includegraphics[height=4cm]{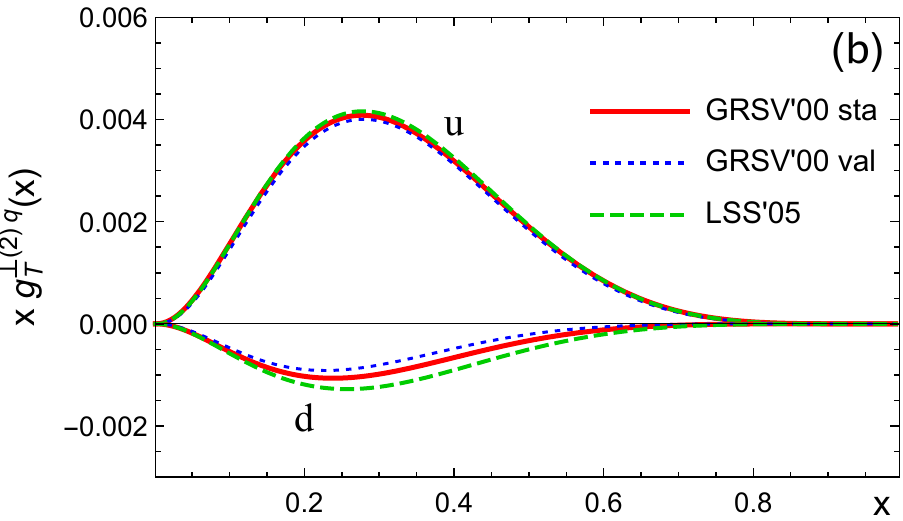}

\vspace{5mm}

  \includegraphics[height=4cm]{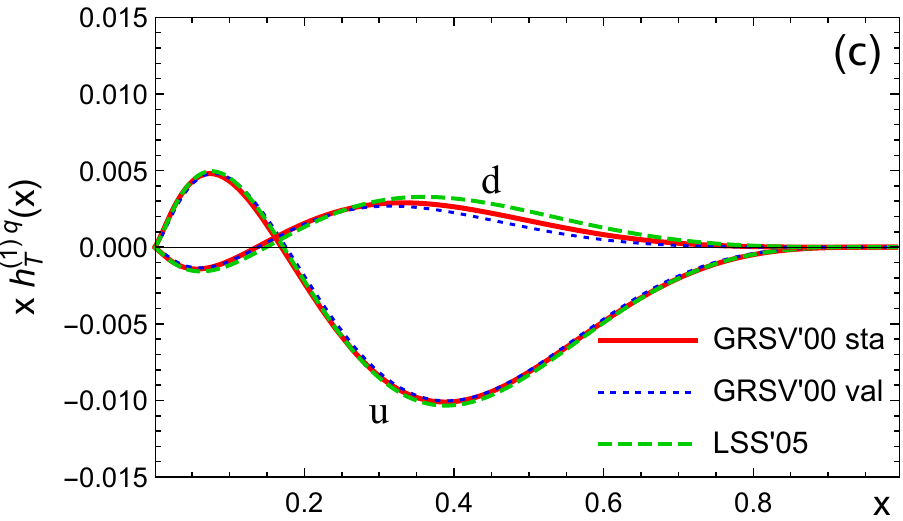} \hspace{1.1cm}
  \includegraphics[height=4cm]{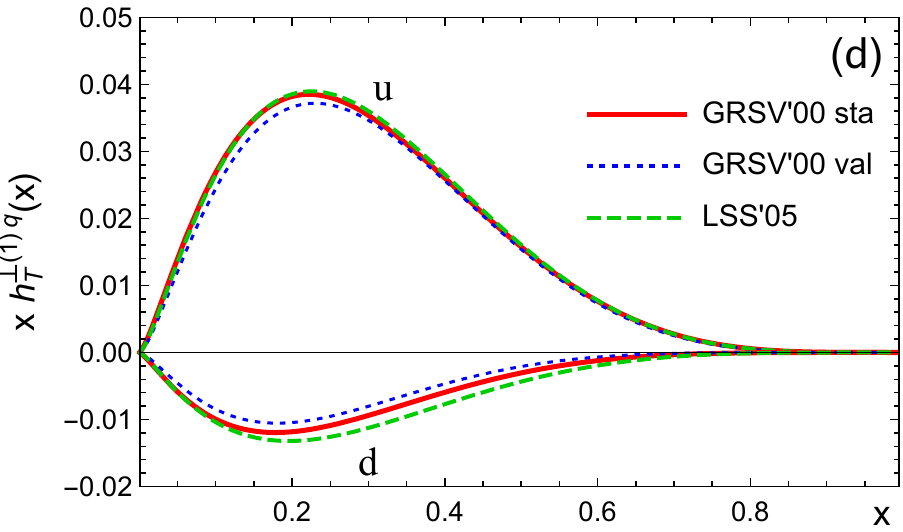}
  \caption{\label{Fig06:glperp-gTperp-hT-hTperp}
Model predictions for the twist-3 polarized TMDs 
(a) $xg_{L}^{\perp(1)q}(x)$, 
(b) $xg_{T}^{\perp(2)q}(x)$, 
(c) $xh_{T}^{(1)q}(x)$, 
(d) $xh_{T}^{\perp(1)q}(x)$ 
as functions of $x$ at the scale $2.5$ GeV$^2$ obtained using the 
helicity LO input PDFs from Fig.~\ref{Fig01:f1-g1}b.}
\end{figure}

Finally, we show the predictions for
$xg_{L}^{\perp(1)q}(x)$ in Fig.~\ref{Fig06:glperp-gTperp-hT-hTperp}a,
$xg_{T}^{\perp(2)q}(x)$ in Fig.~\ref{Fig06:glperp-gTperp-hT-hTperp}b, 
$xh_{T}^{(1)q}(x)$     in Fig.~\ref{Fig06:glperp-gTperp-hT-hTperp}c,
$xh_{T}^{\perp(1)q}(x)$ in Fig.~\ref{Fig06:glperp-gTperp-hT-hTperp}d.
These transverse moments are rather small, and the contributions
of these TMDs to the SIDIS structure functions can be expected to
be small. Currently nothing is known about those TMDs from
phenomenology \cite{Bastami:2018xqd}. The model predicts the
relation $g_{L}^{\perp q}(x,p_T)=-h_{T}^{q}(x,p_T)$  which will be
interesting to test experimentally, and both TMDs exhibit a
node around $x\approx 0.15$ which is also observed in the bag
model \cite{Avakian:2010br}. The same relation between 
$g_{L}^{\perp q}(x,p_T)$ and $h_{T}^{q}(x,p_T)$ holds also in the spectator 
and bag model \cite{Jakob:1997wg,Avakian:2010br} which may hint at a
possible more general underlying quark model symmetry responsible
for such relations among twist-3 TMDs. It would be interesting to
investigate this point in more detail.

\section{Conclusions}
\label{Sec-7:conclusions}

In this work we have generalized the CPM which was originally formulated
to describe PDFs accessible in DIS through an intuitive modeling of the
hadronic tensor. We have shown that the new formulation of the CPM allows
one to reproduce all results for the T-even twist-2 TMDs 
$f_1^q$, $g_1^q$, $h_1^q$, $g_{1T}^{\perp q}$, $h_{1L}^{\perp q}$, $h_{1T}^{\perp q}$
and the twist-3 $g_T^q$ known from prior studies. The advantage of
the new formulation is that it allows one to evaluate systematically 
quark correlators in the CPM. We have demonstrated this by deriving 
the model expressions for all twist-3 T-even TMDs. 

We have checked the consistency of the model by showing that the
QCD equations-of-motion relations are valid in the CPM with 
tilde-terms consistently being zero which is to be expected in a parton 
model approach. The model also complies with Lorentz-invariance 
relations which are valid in quark models which respect 
Lorentz symmetry but lack explicit gauge field degrees of freedom.
We have investigated the relations among TMDs in the CPM. Some of 
these relations were known from prior studies in other models, but 
most of them are specific to our model. The relations among TMDs 
may constitute one of the most interesting predictions of the CPM 
and they will allow one to test the underlying model concepts 
quantitatively in future when more information about TMDs will 
become available. 


We presented numerical predictions for the T-even TMDs and compared 
to studies in other models or lattice QCD, and confronted them with 
TMD parametrizations available for transversity and pretzelosity 
(the latter with very large uncertainties). No extractions are available 
for other TMDs and the model predictions await phenomenological tests
in those cases.
One interesting advantage is that the results from the CPM refer to a 
high renormalization scale where the partonic interpretation may be 
assumed to be valid, while to the best of our knowledge the results 
from other quark model approaches refer to very low hadronic scales 
$\mu < 1\,{\rm GeV}$
\cite{Jaffe:1991ra,Yuan:2003wk,Courtoy:2008vi,Avakian:2008dz,Courtoy:2008dn,Avakian:2010br,Jakob:1997wg,Gamberg:2007wm,Cloet:2007em,Bacchetta:2008af,She:2009jq,Lu:2012gu,Maji:2015vsa,Maji:2016yqo,Maji:2017bcz,Diakonov:1996sr,Diakonov:1997vc,Gamberg:1998vg,Pobylitsa:1998tk,Goeke:2000wv,Wakamatsu:2000fd,Schweitzer:2001sr,Schweitzer:2003uy,Wakamatsu:2003uu,Ohnishi:2003mf,Cebulla:2007ej,Wakamatsu:2009fn,Schweitzer:2012hh,Pasquini:2008ax,Pasquini:2010af,Lorce:2011dv,Boffi:2009sh,Pasquini:2011tk,Lorce:2014hxa,Kofler:2017uzq,Pasquini:2018oyz,Matevosyan:2011vj,Yazdi:2014zaa,Maji:2017wwd,Lyubovitskij:2020otz,Kundu:2001pk,Meissner:2007rx,Mukherjee:2009uy,Mukherjee:2010iw,Xu:2019xhk}.


The new formulation of the CPM may have further interesting applications
going far beyond the computation of the twist-3 TMDs presented in this work.
It would be interesting to introduce a consistent modeling of offshellness
effects. This would allow one to compute tilde-functions and perhaps also
describe T-odd TMDs. Another interesting application could be the
extension of the model to antiquark correlators or gluon correlators.
These aspects will be addressed in future studies.

\ \\
{\bf Acknowledgments.}
We wish to thank E.~Leader,  D.~B.~Stamenov, and W.~Vogelsang for providing
us with the LO parametrizations of the Refs.~\cite{Gluck:2000dy,Leader:2005ci}
and Barbara Pasquini for helpful discussions.
The work of S.B.\ and P.S.\ was supported by the National Science Foundation 
under the Award No.\ 1812423. The work of P.Z.\ was supported by the 
grant LTT17018 of the MEYS (Czech Republic).

\appendix
\section{Model results for TMDs in compact notation}
\label{Appendix}

In order to have a better overview, it is convenient to 
introduce the compact notation for the integration measures
\ba
 \label{eq14}
	&&\{\mathrm{d}p^{\scriptscriptstyle 1}_{{\cal G}q}\} 
        = \frac{\mathrm{d}p^1}{p^0} \, \frac{\mathcal{G}^q(p^0)}{(p^0+m)} \, 
        \delta(x-\frac{p^0+p^1}{M}) \nonumber\\
	&&\{\mathrm{d} p^{\scriptscriptstyle 1}_{{\cal H}q}\} 
        = \frac{\mathrm{d}p^1}{p^0} \, \frac{\mathcal{H}^q(p^0)}{(p^0+m)} \, 
        \delta(x-\frac{p^0+p^1}{M})  \,.
\ea
Then the model results for the leading twist T-even TMDs 
can be summarized as follows
\bsub\label{eq:result-twist-2-SUMMARY}
\begin{align}
	f_1^q(x,p_T) \, = & \, 
         \int \{\mathrm{d}p_{{\cal G}q}^1\} \, \biggl[M x (p^0+m) \biggr]\\
	 g_1^q(x,p_T) \, = & \, 
         \int \{\mathrm{d} p^1_{{\cal H}q}\} \, 
         \biggl[M x (p^0+m) - \vec{p}_T^{\,2}\biggr] \\
	 g_{1T}^{\perp q}(x,p_T) \, = & \, 
         \int \{\mathrm{d} p^1_{{\cal H}q}\} \, \biggl[M (M x + m)\biggr]\\
	 h_1^a(x,p_T) \, = & \, 
         \int \{\mathrm{d} p^1_{{\cal H}q}\} \, 
         \biggl[M x (p^0+m) - \frac{\vec{p}_T^{\,2}}{2}\biggr] 
         \\
	 h_{1L}^{\perp q}(x,p_T) \, = & \, 
         \int \{\mathrm{d} p^1_{{\cal H}q}\} \, \biggl[-M (M x+m) \biggr] \\
	 h_{1T}^{\perp q}(x,p_T) \, = & \, 
         \int \{\mathrm{d} p^1_{{\cal H}q}\} \, \biggl[-M^2 \biggr]\,.
\end{align}
\esub
and the twist-3 T-even TMDs are given by
\bsub\label{eq:result-twist-3-SUMMARY}
\begin{align}
  f^{\perp q}(x,p_T) \, = & \, \int \{\mathrm{d} p^1_{{\cal G}q}\}
  \biggl[M (p^0+m)\biggr] \label{Eq:model-fperp-SUM}\\
  e^q(x,p_T) \, = & \, \int \{\mathrm{d} p^1_{{\cal G}q}\} 
  \biggl[m (p^0+m)\biggr] \label{Eq:model-e-SUM}\\
  g_L^{\perp q}(x,p_T) \, = & \, \int \{\mathrm{d} p^1_{{\cal H}q}\} 
  \biggl[M(Mx - p^0)\biggr] \label{Eq:model-gLperp-SUM}\\
  g_T^q(x,p_T) \, = & \, \int \{\mathrm{d} p^1_{{\cal H}q}\} 
  \biggl[m (p^0+m) + \frac12\vec{p}_T^{\;2}\biggr] \label{Eq:model-gT-SUM}\\
  g_T^{\perp q}(x,p_T) \, = & \, \int \{\mathrm{d} p^1_{{\cal H}q}\} 
  \biggl[M^2 \biggr]\label{Eq:model-gTperp-SUM}\\
  h_L^q(x,p_T) \, = & \, \int \{\mathrm{d} p^1_{{\cal H}q}\} 
  \biggl[m (p^0+m) + \vec{p}_T^{\,2}\biggr] \label{Eq:model-hL-SUM}\\
  h_T^q(x,p_T) \, = & \, \int \{\mathrm{d} p^1_{{\cal H}q}\} 
  \biggl[M (p^0-M x)\biggr] \,. \label{Eq:model-hT-SUM}\\
  h_T^{\perp q}(x,p_T) \, = & \, \int \{\mathrm{d} p^1_{{\cal H}q}\}
  \biggl[M (p^0+m)\biggr] \label{Eq:model-hTperp-SUM}
\end{align}
\esub
The model expressions can be rewritten by exploring 
the onshellness of partons which gives rise to identities like 
$(p^0+p^1)(p^0+m) -\frac12p_T^2=\frac12(xM+m)^2$ and can be used to 
express transversity e.g.\ as
$h_1^q(x,p_T) =\int\{\mathrm{d} p^1_{{\cal H}q}\} \,\frac12(xM+m)^2$.


\begin{thebibliography}{99}

\bibitem{Accardi:2012qut}
  A.~Accardi \textit{et al.},
  Eur. Phys. J. A \textbf{52}, 268 (2016)
  doi:10.1140/epja/i2016-16268-9
  [arXiv:1212.1701 [nucl-ex]].\\
  D.~Boer \textit{et al.},
  [arXiv:1108.1713 [nucl-th]].

\bibitem{Kotzinian:1994dv}
  A.~Kotzinian,
  Nucl. Phys. B \textbf{441}, 234-248 (1995)
  [arXiv:hep-ph/9412283 [hep-ph]].

\bibitem{Mulders:1995dh}
  P.~Mulders and R.~Tangerman,
  Nucl. Phys. B \textbf{461}, 197-237 (1996)
  [arXiv:hep-ph/9510301 [hep-ph]].

\bibitem{Bacchetta:2006tn} 
  A.~Bacchetta, M.~Diehl, K.~Goeke, A.~Metz, P.~J.~Mulders and M.~Schlegel,
  JHEP {\bf 0702}, 093 (2007)
  [hep-ph/0611265].

\bibitem{Metz:2016swz}
  A.~Metz and A.~Vossen,
  Prog. Part. Nucl. Phys. \textbf{91}, 136-202 (2016)
  [arXiv:1607.02521 [hep-ex]].

\bibitem{Boer:1997nt}
  D.~Boer and P.~J.~Mulders,
  Phys. Rev. D \textbf{57}, 5780-5786 (1998)
  [arXiv:hep-ph/9711485 [hep-ph]].

\bibitem{Arnold:2008kf}
  S.~Arnold, A.~Metz and M.~Schlegel,
  Phys. Rev. D \textbf{79}, 034005 (2009)
  [arXiv:0809.2262 [hep-ph]].

\bibitem{Collins:1981uk}
  J.~C.~Collins and D.~E.~Soper,
  Nucl. Phys. B \textbf{193}, 381 (1981)
  [erratum: Nucl. Phys. B \textbf{213}, 545 (1983)].

\bibitem{Efremov:1981sh}
  A.~V.~Efremov and O.~V.~Teryaev,
  Sov. J. Nucl. Phys. \textbf{36}, 140 (1982)
  JINR-P2-81-485.

\bibitem{Efremov:1984ip}
A.~V.~Efremov and O.~V.~Teryaev,
Phys. Lett. B \textbf{150}, 383 (1985).

\bibitem{Collins:1984kg}
  J.~C.~Collins, D.~E.~Soper and G.~F.~Sterman,
  Nucl. Phys. B \textbf{250}, 199-224 (1985).

\bibitem{Qiu:1991pp}
  J.~W.~Qiu and G.~F.~Sterman,
  Phys. Rev. Lett. \textbf{67}, 2264-2267 (1991).

\bibitem{Ji:2004wu}
  X.~D.~Ji, J.~P.~Ma and F.~Yuan,
  Phys. Rev. D \textbf{71}, 034005 (2005)
  [arXiv:hep-ph/0404183 [hep-ph]].

\bibitem{Ji:2006br}
  X.~D.~Ji, J.~W.~Qiu, W.~Vogelsang and F.~Yuan,
  Phys. Lett. B \textbf{638}, 178-186 (2006)
  [arXiv:hep-ph/0604128 [hep-ph]].

\bibitem{Ji:2006vf}
  X.~Ji, J.~w.~Qiu, W.~Vogelsang and F.~Yuan,
  Phys. Rev. D \textbf{73}, 094017 (2006)
  [arXiv:hep-ph/0604023 [hep-ph]].

\bibitem{Collins:2011zzd}
  J.~Collins,
  Camb. Monogr. Part. Phys. Nucl. Phys. Cosmol. \textbf{32}, 1-624 (2011).

\bibitem{Aybat:2011zv}
  S.~M.~Aybat and T.~C.~Rogers,
  Phys. Rev. D \textbf{83}, 114042 (2011)
  [arXiv:1101.5057 [hep-ph]].

\bibitem{Bacchetta:2013pqa}
  A.~Bacchetta and A.~Prokudin,
  Nucl. Phys. B \textbf{875}, 536-551 (2013)
  [arXiv:1303.2129 [hep-ph]].

\bibitem{Sun:2013dya}
  P.~Sun and F.~Yuan,
  Phys. Rev. D \textbf{88}, 034016 (2013)
  [arXiv:1304.5037 [hep-ph]].

\bibitem{Echevarria:2014xaa}
  M.~G.~Echevarria, A.~Idilbi, Z.~B.~Kang and I.~Vitev,
  Phys. Rev. D \textbf{89}, 074013 (2014)
  [arXiv:1401.5078 [hep-ph]].

\bibitem{Collins:2014jpa}
  J.~Collins and T.~Rogers,
  Phys. Rev. D \textbf{91}, 074020 (2015)
  [arXiv:1412.3820 [hep-ph]].

\bibitem{Collins:2016hqq}
  J.~Collins, L.~Gamberg, A.~Prokudin, T.~C.~Rogers, N.~Sato and B.~Wang,
  Phys. Rev. D \textbf{94}, 034014 (2016)
  [arXiv:1605.00671 [hep-ph]].

\bibitem{Gehrmann:2014yya}
  T.~Gehrmann, T.~Luebbert and L.~L.~Yang,
  JHEP \textbf{06}, 155 (2014)
  [arXiv:1403.6451 [hep-ph]].

\bibitem{Echevarria:2015byo}
  M.~G.~Echevarria, I.~Scimemi and A.~Vladimirov,
  Phys. Rev. D \textbf{93}, 054004 (2016)
  [arXiv:1511.05590 [hep-ph]].

\bibitem{Echevarria:2016scs}
  M.~G.~Echevarria, I.~Scimemi and A.~Vladimirov,
  JHEP \textbf{09}, 004 (2016)
  [arXiv:1604.07869 [hep-ph]].

\bibitem{Li:2016ctv}
  Y.~Li and H.~X.~Zhu,
  Phys. Rev. Lett. \textbf{118}, 022004 (2017)
  [arXiv:1604.01404 [hep-ph]].

\bibitem{Vladimirov:2016dll}
  A.~A.~Vladimirov,
  Phys. Rev. Lett. \textbf{118}, 062001 (2017)
  [arXiv:1610.05791 [hep-ph]].

\bibitem{Gutierrez-Reyes:2017glx}
  D.~Guti\'errez-Reyes, I.~Scimemi and A.~A.~Vladimirov,
  Phys. Lett. B \textbf{769}, 84-89 (2017)
  [arXiv:1702.06558 [hep-ph]].

\bibitem{Gutierrez-Reyes:2018iod}
  D.~Gutierrez-Reyes, I.~Scimemi and A.~Vladimirov,
  JHEP \textbf{07}, 172 (2018)
  [arXiv:1805.07243 [hep-ph]].

\bibitem{Luo:2019hmp}
  M.~X.~Luo, X.~Wang, X.~Xu, L.~L.~Yang, T.~Z.~Yang and H.~X.~Zhu,
  JHEP \textbf{10}, 083 (2019)
  [arXiv:1908.03831 [hep-ph]].

\bibitem{Luo:2019szz}
  M.~X.~Luo, T.~Z.~Yang, H.~X.~Zhu and Y.~J.~Zhu,
  Phys. Rev. Lett. \textbf{124}, 092001 (2020)
  [arXiv:1912.05778 [hep-ph]].

\bibitem{Ebert:2020yqt}
  M.~A.~Ebert, B.~Mistlberger and G.~Vita,
  JHEP \textbf{09}, 146 (2020)
  [arXiv:2006.05329 [hep-ph]].

\bibitem{Efremov:2004tp}
  A.~V.~Efremov, K.~Goeke, S.~Menzel, A.~Metz and P.~Schweitzer,
  Phys. Lett. B \textbf{612}, 233-244 (2005)
  [arXiv:hep-ph/0412353 [hep-ph]].

\bibitem{Anselmino:2005nn}
  M.~Anselmino, M.~Boglione, U.~D'Alesio, A.~Kotzinian, F.~Murgia 
  and A.~Prokudin,
  Phys. Rev. D \textbf{71}, 074006 (2005)
  [arXiv:hep-ph/0501196 [hep-ph]].

\bibitem{Vogelsang:2005cs}
  W.~Vogelsang and F.~Yuan,
  Phys. Rev. D \textbf{72}, 054028 (2005)
  [arXiv:hep-ph/0507266 [hep-ph]].

\bibitem{Collins:2005ie}
  J.~C.~Collins, A.~V.~Efremov, K.~Goeke, S.~Menzel, A.~Metz and P.~Schweitzer,
  Phys. Rev. D \textbf{73}, 014021 (2006)
  [arXiv:hep-ph/0509076 [hep-ph]].

\bibitem{Collins:2005rq}
  J.~C.~Collins, A.~V.~Efremov, K.~Goeke, M.~Grosse Perdekamp, S.~Menzel, B.~Meredith, A.~Metz and P.~Schweitzer,
  Phys. Rev. D \textbf{73}, 094023 (2006)
  [arXiv:hep-ph/0511272 [hep-ph]].

\bibitem{Anselmino:2007fs}
  M.~Anselmino, M.~Boglione, U.~D'Alesio, A.~Kotzinian, F.~Murgia, 
  A.~Prokudin and C.~Turk,
  Phys. Rev. D \textbf{75}, 054032 (2007)
  [arXiv:hep-ph/0701006 [hep-ph]].

\bibitem{Anselmino:2013vqa}
  M.~Anselmino, M.~Boglione, U.~D'Alesio, S.~Melis, F.~Murgia and A.~Prokudin,
  Phys. Rev. D \textbf{87}, 094019 (2013)
  [arXiv:1303.3822 [hep-ph]].

\bibitem{Signori:2013mda}
  A.~Signori, A.~Bacchetta, M.~Radici and G.~Schnell,
  JHEP \textbf{11}, 194 (2013)
  [arXiv:1309.3507 [hep-ph]].

\bibitem{Anselmino:2013lza}
  M.~Anselmino, M.~Boglione, J.~O.~Gonzalez Hernandez, S.~Melis and A.~Prokudin,
  JHEP \textbf{04}, 005 (2014)
  [arXiv:1312.6261 [hep-ph]].

\bibitem{Kang:2014zza}
  Z.~B.~Kang, A.~Prokudin, P.~Sun and F.~Yuan,
  Phys. Rev. D \textbf{91}, 071501 (2015)
  [arXiv:1410.4877 [hep-ph]].

\bibitem{Kang:2015msa}
  Z.~B.~Kang, A.~Prokudin, P.~Sun and F.~Yuan,
  Phys. Rev. D \textbf{93},  014009 (2016)
  [arXiv:1505.05589 [hep-ph]].

\bibitem{Kang:2017btw}
  Z.~B.~Kang, A.~Prokudin, F.~Ringer and F.~Yuan,
  Phys. Lett. B \textbf{774}, 635-642 (2017)
  [arXiv:1707.00913 [hep-ph]].

\bibitem{Cammarota:2020qcw}
  J.~Cammarota \textit{et al.} [Jefferson Lab Angular Momentum],
  Phys. Rev. D \textbf{102}, 054002 (2020)
  [arXiv:2002.08384 [hep-ph]].

\bibitem{Lefky:2014eia}
  C.~Lefky and A.~Prokudin,
  Phys. Rev. D \textbf{91}, 034010 (2015)
  [arXiv:1411.0580 [hep-ph]].

\bibitem{Collins:2003fm}
  J.~C.~Collins,
  Acta Phys. Polon. B \textbf{34}, 3103 (2003)
  [arXiv:hep-ph/0304122 [hep-ph]].

\bibitem{DAlesio:2007bjf}
  U.~D'Alesio and F.~Murgia,
  Prog. Part. Nucl. Phys. \textbf{61}, 394-454 (2008)
  [arXiv:0712.4328 [hep-ph]].

\bibitem{Barone:2010zz}
  V.~Barone, F.~Bradamante and A.~Martin,
  Prog. Part. Nucl. Phys. \textbf{65}, 267-333 (2010)
  [arXiv:1011.0909 [hep-ph]].

\bibitem{Aidala:2012mv}
  C.~A.~Aidala, S.~D.~Bass, D.~Hasch and G.~K.~Mallot,
  Rev. Mod. Phys. \textbf{85}, 655-691 (2013)
  [arXiv:1209.2803 [hep-ph]].

\bibitem{Avakian:2019drf}
  H.~Avakian, B.~Parsamyan and A.~Prokudin,
  Riv. Nuovo Cim. \textbf{42}, 1-48 (2019)
  [arXiv:1909.13664 [hep-ex]].

\bibitem{Anselmino:2020vlp}
  M.~Anselmino, A.~Mukherjee and A.~Vossen,
  Prog. Part. Nucl. Phys. \textbf{114}, 103806 (2020)
  doi:10.1016/j.ppnp.2020.103806
  [arXiv:2001.05415 [hep-ph]].

\bibitem{Zavada:1996kp} 
  P.~Zavada,
  Phys.\ Rev.\ D {\bf 55}, 4290 (1997)
  [hep-ph/9609372].

\bibitem{Zavada:2001bq} 
  P.~Zavada,
  Phys.\ Rev.\ D {\bf 65}, 054040 (2002)
  [hep-ph/0106215].

\bibitem{Zavada:2002uz} 
  P.~Zavada,
  Phys.\ Rev.\ D {\bf 67}, 014019 (2003)
  [hep-ph/0210141].

\bibitem{Efremov:2004tz} 
  A.~V.~Efremov, O.~V.~Teryaev and P.~Zavada,
  Phys.\ Rev.\ D {\bf 70}, 054018 (2004)
  [hep-ph/0405225].

\bibitem{Zavada:2007ww} 
  P.~Zavada,
  Eur.\ Phys.\ J.\ C {\bf 52}, 121 (2007)
  [arXiv:0706.2988 [hep-ph]].

\bibitem{Efremov:2009ze} 
  A.~V.~Efremov, P.~Schweitzer, O.~V.~Teryaev and P.~Zavada,
  Phys.\ Rev.\ D {\bf 80}, 014021 (2009)
  [arXiv:0903.3490 [hep-ph]].

\bibitem{Zavada:2009ska} 
  P.~Zavada,
  Phys.\ Rev.\ D {\bf 83}, 014022 (2011)
  [arXiv:0908.2316 [hep-ph]].

\bibitem{Avakian:2009jt} 
  H.~Avakian, A.~V.~Efremov, P.~Schweitzer, O.~V.~Teryaev, F.~Yuan and P.~Zavada,
  Mod.\ Phys.\ Lett.\ A {\bf 24}, 2995 (2009)
  [arXiv:0910.3181 [hep-ph]].

\bibitem{Efremov:2010mt} 
  A.~V.~Efremov, P.~Schweitzer, O.~V.~Teryaev and P.~Zavada,
  Phys.\ Rev.\ D {\bf 83}, 054025 (2011)
  [arXiv:1012.5296 [hep-ph]].

\bibitem{Zavada:2011cv} 
  P.~Zavada,
  Phys.\ Rev.\ D {\bf 85}, 037501 (2012)
  [arXiv:1106.5607 [hep-ph]].

\bibitem{Zavada:2013ola} 
  P.~Zavada,
  Phys.\ Rev.\ D {\bf 89}, 014012 (2014)
  [arXiv:1307.0699 [hep-ph]].

\bibitem{Zavada:2015gaa} 
  P.~Zavada,
  Phys.\ Lett.\ B {\bf 751}, 525 (2015)
  [arXiv:1503.07924 [hep-ph]].

\bibitem{Zavada:2019yom}
  P.~Zavada,
  [arXiv:1911.12703 [hep-ph]].

\bibitem{Feynman:1973xc}
  R.~P.~Feynman,
  ``Photon-hadron interactions,''
  (Reading, 1972).

\bibitem{Jaffe:1991ra}
  R.~L.~Jaffe and X.~D.~Ji,
  Nucl. Phys. B \textbf{375}, 527-560 (1992).

\bibitem{Yuan:2003wk}
  F.~Yuan,
  Phys. Lett. B \textbf{575}, 45-54 (2003)
  [arXiv:hep-ph/0308157 [hep-ph]].

\bibitem{Courtoy:2008vi}
  A.~Courtoy, F.~Fratini, S.~Scopetta and V.~Vento,
  Phys. Rev. D \textbf{78}, 034002 (2008)
  [arXiv:0801.4347 [hep-ph]].

\bibitem{Avakian:2008dz}
  H.~Avakian, A.~V.~Efremov, P.~Schweitzer and F.~Yuan,
  Phys. Rev. D \textbf{78}, 114024 (2008)
  [arXiv:0805.3355 [hep-ph]].

\bibitem{Courtoy:2008dn}
  A.~Courtoy, S.~Scopetta and V.~Vento,
  Phys. Rev. D \textbf{79}, 074001 (2009)
  [arXiv:0811.1191 [hep-ph]].

\bibitem{Avakian:2010br}
  H.~Avakian, A.~Efremov, P.~Schweitzer and F.~Yuan,
  Phys. Rev. D \textbf{81}, 074035 (2010)
  [arXiv:1001.5467 [hep-ph]].

\bibitem{Jakob:1997wg}
  R.~Jakob, P.~J.~Mulders and J.~Rodrigues,
  Nucl. Phys. A \textbf{626}, 937-965 (1997)
  [arXiv:hep-ph/9704335 [hep-ph]].

\bibitem{Gamberg:2007wm}
  L.~P.~Gamberg, G.~R.~Goldstein and M.~Schlegel,
  Phys. Rev. D \textbf{77}, 094016 (2008)
  [arXiv:0708.0324 [hep-ph]].

\bibitem{Cloet:2007em}
  I.~C.~Cloet, W.~Bentz and A.~W.~Thomas,
  Phys. Lett. B \textbf{659}, 214-220 (2008)
  [arXiv:0708.3246 [hep-ph]].

\bibitem{Bacchetta:2008af}
  A.~Bacchetta, F.~Conti and M.~Radici,
  Phys. Rev. D \textbf{78}, 074010 (2008)
  [arXiv:0807.0323 [hep-ph]].

\bibitem{She:2009jq}
  J.~She, J.~Zhu and B.~Q.~Ma,
  Phys. Rev. D \textbf{79}, 054008 (2009)
  [arXiv:0902.3718 [hep-ph]].

\bibitem{Lu:2012gu}
  Z.~Lu and I.~Schmidt,
  Phys. Lett. B \textbf{712}, 451-455 (2012)
  [arXiv:1202.0700 [hep-ph]].

\bibitem{Maji:2015vsa}
  T.~Maji, C.~Mondal, D.~Chakrabarti and O.~V.~Teryaev,
  JHEP \textbf{01}, 165 (2016)
  [arXiv:1506.04560 [hep-ph]].

\bibitem{Maji:2016yqo}
  T.~Maji and D.~Chakrabarti,
  Phys. Rev. D \textbf{94}, 094020 (2016)
  [arXiv:1608.07776 [hep-ph]].

\bibitem{Maji:2017bcz}
  T.~Maji and D.~Chakrabarti,
  Phys. Rev. D \textbf{95}, 074009 (2017)
  [arXiv:1702.04557 [hep-ph]].

\bibitem{Diakonov:1996sr}
  D.~Diakonov, V.~Petrov, P.~Pobylitsa, M.~V.~Polyakov and C.~Weiss,
  Nucl.\ Phys.\  B {\bf 480} (1996) 341
  [arXiv:hep-ph/9606314].

\bibitem{Diakonov:1997vc}
  D.~Diakonov, V.~Petrov, P.~Pobylitsa, M.~Polyakov and C.~Weiss,
  Phys.\ Rev.\ D {\bf 56}, 4069 (1997)
  [arXiv:hep-ph/9703420].

\bibitem{Gamberg:1998vg}
  L.~P.~Gamberg, H.~Reinhardt and H.~Weigel,
  Phys. Rev. D \textbf{58}, 054014 (1998)
  [arXiv:hep-ph/9801379 [hep-ph]].

\bibitem{Pobylitsa:1998tk} 
  P.~V.~Pobylitsa, M.~V.~Polyakov, K.~Goeke, T.~Watabe and C.~Weiss,
  Phys.\ Rev.\ D {\bf 59}, 034024 (1999)
  [hep-ph/9804436].

\bibitem{Goeke:2000wv}
  K.~Goeke, P.~V.~Pobylitsa, M.~V.~Polyakov, P.~Schweitzer and D.~Urbano,
  Acta Phys. Polon. B \textbf{32}, 1201-1224 (2001)
  [arXiv:hep-ph/0001272 [hep-ph]].

\bibitem{Wakamatsu:2000fd}
  M.~Wakamatsu,
  Phys. Lett. B \textbf{509}, 59-68 (2001)
  [arXiv:hep-ph/0012331 [hep-ph]].

\bibitem{Schweitzer:2001sr}
  P.~Schweitzer, D.~Urbano, M.~V.~Polyakov, C.~Weiss, P.~V.~Pobylitsa 
  and K.~Goeke,
  Phys. Rev. D \textbf{64}, 034013 (2001)
  [arXiv:hep-ph/0101300 [hep-ph]].

\bibitem{Schweitzer:2003uy}
  P.~Schweitzer,
  Phys. Rev. D \textbf{67}, 114010 (2003)
  [arXiv:hep-ph/0303011 [hep-ph]].

\bibitem{Wakamatsu:2003uu}
  M.~Wakamatsu and Y.~Ohnishi,
  Phys. Rev. D \textbf{67}, 114011 (2003)
  [arXiv:hep-ph/0303007 [hep-ph]].

\bibitem{Ohnishi:2003mf}
  Y.~Ohnishi and M.~Wakamatsu,
  Phys. Rev. D \textbf{69}, 114002 (2004)
  [arXiv:hep-ph/0312044 [hep-ph]].

\bibitem{Cebulla:2007ej}
  C.~Cebulla, J.~Ossmann, P.~Schweitzer and D.~Urbano,
  Acta Phys. Polon. B \textbf{39}, 609-640 (2008)
  [arXiv:0710.3103 [hep-ph]].

\bibitem{Wakamatsu:2009fn}
  M.~Wakamatsu,
  Phys. Rev. D \textbf{79}, 094028 (2009)
  [arXiv:0903.1886 [hep-ph]].

\bibitem{Schweitzer:2012hh}
  P.~Schweitzer, M.~Strikman and C.~Weiss,
  JHEP \textbf{01}, 163 (2013)
  [arXiv:1210.1267 [hep-ph]].

\bibitem{Pasquini:2008ax}
  B.~Pasquini, S.~Cazzaniga and S.~Boffi,
  Phys. Rev. D \textbf{78}, 034025 (2008)
  [arXiv:0806.2298 [hep-ph]].

\bibitem{Pasquini:2010af}
  B.~Pasquini and F.~Yuan,
  Phys. Rev. D \textbf{81}, 114013 (2010)
  [arXiv:1001.5398 [hep-ph]].

\bibitem{Boffi:2009sh}
  S.~Boffi, A.~V.~Efremov, B.~Pasquini and P.~Schweitzer,
  Phys. Rev. D \textbf{79}, 094012 (2009)
  [arXiv:0903.1271 [hep-ph]].

\bibitem{Lorce:2011dv}
  C.~Lorc\'e, B.~Pasquini and M.~Vanderhaeghen,
  JHEP \textbf{05}, 041 (2011)
  [arXiv:1102.4704 [hep-ph]].

\bibitem{Pasquini:2011tk}
  B.~Pasquini and P.~Schweitzer,
  Phys. Rev. D \textbf{83}, 114044 (2011)
  [arXiv:1103.5977 [hep-ph]].

\bibitem{Lorce:2014hxa}
  C.~Lorc\'e, B.~Pasquini and P.~Schweitzer,
  JHEP \textbf{01}, 103 (2015)
  [arXiv:1411.2550 [hep-ph]].

\bibitem{Kofler:2017uzq}
  S.~Kofler and B.~Pasquini,
  Phys. Rev. D \textbf{95}, 094015 (2017)
  [arXiv:1701.07839 [hep-ph]].

\bibitem{Pasquini:2018oyz}
  B.~Pasquini and S.~Rodini,
  Phys. Lett. B \textbf{788}, 414-424 (2019)
  [arXiv:1806.10932 [hep-ph]].

\bibitem{Matevosyan:2011vj}
  H.~H.~Matevosyan, W.~Bentz, I.~C.~Cloet and A.~W.~Thomas,
  Phys. Rev. D \textbf{85}, 014021 (2012)
  [arXiv:1111.1740 [hep-ph]].

\bibitem{Yazdi:2014zaa}
  Z.~Alizadeh Yazdi, F.~Taghavi-Shahri, F.~Arash and M.~E.~Zomorrodian,
  Phys. Rev. C \textbf{89}, 055201 (2014)
  [arXiv:1401.1295 [hep-ph]].

\bibitem{Maji:2017wwd}
  T.~Maji, D.~Chakrabarti and A.~Mukherjee,
  Phys. Rev. D \textbf{97}, 014016 (2018)
  [arXiv:1711.02930 [hep-ph]].

\bibitem{Lyubovitskij:2020otz}
  V.~E.~Lyubovitskij and I.~Schmidt,
  Phys. Rev. D \textbf{102}, 034011 (2020)
  [arXiv:2005.10163 [hep-ph]].

\bibitem{Kundu:2001pk}
  R.~Kundu and A.~Metz,
  Phys. Rev. D \textbf{65}, 014009 (2002)
  [arXiv:hep-ph/0107073 [hep-ph]].

\bibitem{Meissner:2007rx}
  S.~Meissner, A.~Metz and K.~Goeke,
  Phys. Rev. D \textbf{76}, 034002 (2007)
  [arXiv:hep-ph/0703176 [hep-ph]].

\bibitem{Mukherjee:2009uy}
  A.~Mukherjee,
  Phys. Lett. B \textbf{687}, 180-183 (2010)
  [arXiv:0912.1446 [hep-ph]].

\bibitem{Mukherjee:2010iw}
  A.~Mukherjee and R.~Korrapati,
  Mod. Phys. Lett. A \textbf{26}, 2653-2662 (2011)
  [arXiv:1005.2830 [hep-ph]].

\bibitem{Xu:2019xhk}
  C.~Mondal, S.~Xu, J.~Lan, X.~Zhao, Y.~Li, D.~Chakrabarti and J.~P.~Vary,
  Phys. Rev. D \textbf{102}, 016008 (2020)
  [arXiv:1911.10913 [hep-ph]].

\bibitem{Hagler:2009mb}
  P.~H\"agler, B.~U.~Musch, J.~W.~Negele and A.~Sch\"afer,
  EPL \textbf{88}, 61001 (2009)
  [arXiv:0908.1283 [hep-lat]].

\bibitem{Musch:2010ka}
  B.~U.~Musch, P.~H\"agler, J.~W.~Negele and A.~Sch\"afer,
  Phys. Rev. D \textbf{83}, 094507 (2011)
  [arXiv:1011.1213 [hep-lat]].

\bibitem{Musch:2011er}
  B.~U.~Musch, P.~H\"agler, M.~Engelhardt, J.~W.~Negele and A.~Sch\"afer,
  Phys. Rev. D \textbf{85}, 094510 (2012)
  [arXiv:1111.4249 [hep-lat]].

\bibitem{Chen:2016utp}
  J.~W.~Chen, S.~D.~Cohen, X.~Ji, H.~W.~Lin and J.~H.~Zhang,
  Nucl. Phys. B \textbf{911}, 246-273 (2016)
  [arXiv:1603.06664 [hep-ph]].

\bibitem{Alexandrou:2016jqi}
  C.~Alexandrou, K.~Cichy, M.~Constantinou, K.~Hadjiyiannakou, 
  K.~Jansen, F.~Steffens and C.~Wiese,
  Phys. Rev. D \textbf{96}, 014513 (2017)
  [arXiv:1610.03689 [hep-lat]].

\bibitem{Yoon:2017qzo}
  B.~Yoon, M.~Engelhardt, R.~Gupta, T.~Bhattacharya, J.~R.~Green, B.~U.~Musch, 
  J.~W.~Negele, A.~V.~Pochinsky, A.~Sch\"afer and S.~N.~Syritsyn,
  Phys. Rev. D \textbf{96}, 094508 (2017)
  [arXiv:1706.03406 [hep-lat]].

\bibitem{Orginos:2017kos}
  K.~Orginos, A.~Radyushkin, J.~Karpie and S.~Zafeiropoulos,
  Phys. Rev. D \textbf{96}, 094503 (2017)
  [arXiv:1706.05373 [hep-ph]].

\bibitem{Joo:2019jct}
  B.~Jo\'o, J.~Karpie, K.~Orginos, A.~Radyushkin, D.~Richards and
  S.~Zafeiropoulos,
  JHEP \textbf{12}, 081 (2019)
  [arXiv:1908.09771 [hep-lat]].

\bibitem{Lorce:2011zta}
  C.~Lorc\'e and B.~Pasquini,
  Phys. Rev. D \textbf{84}, 034039 (2011)
  [arXiv:1104.5651 [hep-ph]].

\bibitem{Lorce:2011kn}
  C.~Lorce\' and B.~Pasquini,
  Phys. Lett. B \textbf{710}, 486-488 (2012)
  [arXiv:1111.6069 [hep-ph]].

\bibitem{Collins:2002kn}
  J.~C.~Collins,
  Phys. Lett. B \textbf{536}, 43-48 (2002)
  [arXiv:hep-ph/0204004 [hep-ph]].

\bibitem{Brodsky:2002rv}
  S.~J.~Brodsky, D.~S.~Hwang and I.~Schmidt,
  Nucl. Phys. B \textbf{642}, 344-356 (2002)
  [arXiv:hep-ph/0206259 [hep-ph]].

\bibitem{Brodsky:2002cx}
  S.~J.~Brodsky, D.~S.~Hwang and I.~Schmidt,
  Phys. Lett. B \textbf{530}, 99-107 (2002)
  [arXiv:hep-ph/0201296 [hep-ph]].

\bibitem{Belitsky:2002sm}
  A.~V.~Belitsky, X.~Ji and F.~Yuan,
  Nucl. Phys. B \textbf{656}, 165-198 (2003)
  [arXiv:hep-ph/0208038 [hep-ph]].

\bibitem{Bomhof:2006dp}
  C.~J.~Bomhof, P.~J.~Mulders and F.~Pijlman,
  Eur. Phys. J. C \textbf{47}, 147-162 (2006)
  [arXiv:hep-ph/0601171 [hep-ph]].

\bibitem{Goeke:2005hb}
  K.~Goeke, A.~Metz and M.~Schlegel,
  Phys. Lett. B \textbf{618}, 90-96 (2005)
  [arXiv:hep-ph/0504130 [hep-ph]].

\bibitem{Metz:2008ib}
  A.~Metz, P.~Schweitzer and T.~Teckentrup,
  Phys. Lett. B \textbf{680}, 141-147 (2009)
  [arXiv:0810.5212 [hep-ph]].

\bibitem{Pobylitsa:2003ty} 
  P.~V.~Pobylitsa,
  hep-ph/0301236.

\bibitem{Goeke:2003az}
  K.~Goeke, A.~Metz, P.~V.~Pobylitsa and M.~V.~Polyakov,
  Phys. Lett. B \textbf{567}, 27-30 (2003)
  [arXiv:hep-ph/0302028 [hep-ph]].

{}


\bibitem{Karl:1984cz}
  G.~Karl and J.~E.~Paton,
  Phys.\ Rev.\  D {\bf 30}, 238 (1984).

\bibitem{Wandzura:1977qf}
  S.~Wandzura and F.~Wilczek,
  Phys. Lett. B \textbf{72}, 195-198 (1977).

\bibitem{Balla:1997hf}
  J.~Balla, M.~V.~Polyakov and C.~Weiss,
  Nucl. Phys. B \textbf{510}, 327-364 (1998)
  [arXiv:hep-ph/9707515 [hep-ph]].

\bibitem{Dressler:1999hc}
  B.~Dressler and M.~V.~Polyakov,
  Phys. Rev. D \textbf{61}, 097501 (2000)
  [arXiv:hep-ph/9912376 [hep-ph]].

\bibitem{Gockeler:2000ja}
  M.~Gockeler, R.~Horsley, W.~Kurzinger, H.~Oelrich, D.~Pleiter, 
  P.~E.~L.~Rakow, A.~Schafer and G.~Schierholz,
  Phys. Rev. D \textbf{63}, 074506 (2001)
  [arXiv:hep-lat/0011091 [hep-lat]].

\bibitem{Gockeler:2005vw}
  M.~Gockeler, R.~Horsley, D.~Pleiter, P.~E.~L.~Rakow, A.~Schafer, 
  G.~Schierholz, H.~Stuben and J.~M.~Zanotti,
  Phys. Rev. D \textbf{72}, 054507 (2005)
  [arXiv:hep-lat/0506017 [hep-lat]].

\bibitem{Abe:1998wq}
  K.~Abe \textit{et al.} [E143],
  Phys. Rev. D \textbf{58}, 112003 (1998)
  [arXiv:hep-ph/9802357 [hep-ph]].

\bibitem{Anthony:2002hy}
  P.~L.~Anthony \textit{et al.} [E155],
  Phys. Lett. B \textbf{553}, 18-24 (2003)
  [arXiv:hep-ex/0204028 [hep-ex]].

\bibitem{Airapetian:2011wu}
  A.~Airapetian \textit{et al.} [HERMES],
  Eur. Phys. J. C \textbf{72}, 1921 (2012)
  [arXiv:1112.5584 [hep-ex]].

\bibitem{Efremov:2001cz}
  A.~V.~Efremov, K.~Goeke and P.~Schweitzer,
  Phys. Lett. B \textbf{522}, 37-48 (2001)
  [arXiv:hep-ph/0108213 [hep-ph]].

\bibitem{Efremov:2001ia}
  A.~V.~Efremov, K.~Goeke and P.~Schweitzer,
  Eur. Phys. J. C \textbf{24}, 407-412 (2002)
  [arXiv:hep-ph/0112166 [hep-ph]].

\bibitem{Kotzinian:2006dw}
  A.~Kotzinian, B.~Parsamyan and A.~Prokudin,
  Phys. Rev. D \textbf{73}, 114017 (2006)
  [arXiv:hep-ph/0603194 [hep-ph]].

\bibitem{Avakian:2007mv}
  H.~Avakian, A.~V.~Efremov, K.~Goeke, A.~Metz, P.~Schweitzer and T.~Teckentrup,
  Phys. Rev. D \textbf{77}, 014023 (2008)
  [arXiv:0709.3253 [hep-ph]].

\bibitem{Accardi:2009au}
  A.~Accardi, A.~Bacchetta, W.~Melnitchouk and M.~Schlegel,
  JHEP \textbf{11}, 093 (2009)
  [arXiv:0907.2942 [hep-ph]].

\bibitem{Bastami:2018xqd}
  S.~Bastami, H.~Avakian, A.~V.~Efremov, A.~Kotzinian, B.~U.~Musch, 
  B.~Parsamyan, A.~Prokudin, M.~Schlegel, G.~Schnell, P.~Schweitzer 
  and K.~Tezgin,
  JHEP \textbf{06}, 007 (2019)
  [arXiv:1807.10606 [hep-ph]].

\bibitem{Bastami:2020asv}
  S.~Bastami, L.~Gamberg, B.~Parsamyan, B.~Pasquini, 
  A.~Prokudin and P.~Schweitzer,
  [arXiv:2005.14322 [hep-ph]].

\bibitem{Koike:2008du}
  Y.~Koike, K.~Tanaka and S.~Yoshida,
  Phys. Lett. B \textbf{668}, 286-292 (2008)
  [arXiv:0805.2289 [hep-ph]].


%

\bibitem{Brodsky:2002ue}
S.~J.~Brodsky, P.~Hoyer, N.~Marchal, S.~Peigne and F.~Sannino,
Phys. Rev. D \textbf{65}, 114025 (2002)
[arXiv:hep-ph/0104291 [hep-ph]].

\bibitem{Gluck:1998xa}
  M.~Gl\"uck, E.~Reya and A.~Vogt,
  Eur. Phys. J. C \textbf{5}, 461-470 (1998)
  [arXiv:hep-ph/9806404 [hep-ph]].

\bibitem{Martin:1998sq}
  A.~D.~Martin, R.~G.~Roberts, W.~J.~Stirling and R.~S.~Thorne,
  Eur. Phys. J. C \textbf{4}, 463-496 (1998)
  [arXiv:hep-ph/9803445 [hep-ph]].

\bibitem{Martin:2009iq}
  A.~D.~Martin, W.~J.~Stirling, R.~S.~Thorne and G.~Watt,
  Eur. Phys. J. C \textbf{63}, 189-285 (2009)
  [arXiv:0901.0002 [hep-ph]].

\bibitem{Gluck:2000dy}
  M.~Gl\"uck, E.~Reya, M.~Stratmann and W.~Vogelsang,
  Phys. Rev. D \textbf{63}, 094005 (2001)
  [arXiv:hep-ph/0011215 [hep-ph]].

\bibitem{Leader:2005ci}
  E.~Leader, A.~V.~Sidorov and D.~B.~Stamenov,
  Phys. Rev. D \textbf{73}, 034023 (2006)
  [arXiv:hep-ph/0512114 [hep-ph]].

\bibitem{Harland-Lang:2014zoa}
  L.~A.~Harland-Lang, A.~D.~Martin, P.~Motylinski, R.~S.~Thorne,
  Eur. Phys. J. C \textbf{75}, 204 (2015)
  [arXiv:1412.3989 [hep-ph]].

\bibitem{Alekhin:2017kpj}
  S.~Alekhin, J.~Bl\"umlein, S.~Moch and R.~Placakyte,
  Phys. Rev. D \textbf{96}, 014011 (2017)
  [arXiv:1701.05838 [hep-ph]].

\bibitem{Ball:2017nwa}
  R.~D.~Ball \textit{et al.} [NNPDF],
  Eur. Phys. J. C \textbf{77}, 663 (2017)
  [arXiv:1706.00428 [hep-ph]].

\bibitem{Sato:2019yez}
  N.~Sato \textit{et al.} [JAM],
  Phys. Rev. D \textbf{101}, 074020 (2020)
  [arXiv:1905.03788 [hep-ph]].

\bibitem{Hou:2019efy}
  T.~J.~Hou, J.~Gao, T.~J.~Hobbs, K.~Xie, S.~Dulat, M.~Guzzi, J.~Huston, 
  P.~Nadolsky, J.~Pumplin, C.~Schmidt, I.~Sitiwaldi, D.~Stump and C.~P.~Yuan,
  [arXiv:1912.10053 [hep-ph]].

\bibitem{deFlorian:2014yva}
  D.~de Florian, R.~Sassot, M.~Stratmann and W.~Vogelsang,
  Phys. Rev. Lett. \textbf{113}, 012001 (2014)
  [arXiv:1404.4293 [hep-ph]].

\bibitem{Nocera:2014gqa}
  E.~R.~Nocera \textit{et al.} [NNPDF],
  Nucl. Phys. B \textbf{887}, 276 (2014)
  [arXiv:1406.5539 [hep-ph]].

\bibitem{Leader:2014uua}
  E.~Leader, A.~V.~Sidorov and D.~B.~Stamenov,
  Phys. Rev. D \textbf{91}, 054017 (2015)
  [arXiv:1410.1657 [hep-ph]].

\bibitem{Ethier:2017zbq}
  J.~J.~Ethier, N.~Sato and W.~Melnitchouk,
  Phys. Rev. Lett. \textbf{119}, 132001 (2017)
  [arXiv:1705.05889 [hep-ph]].

\bibitem{deFlorian:2019zkl}
  D.~De Florian, G.~A.~Lucero, R.~Sassot, M.~Stratmann and W.~Vogelsang,
  Phys. Rev. D \textbf{100}, 114027 (2019)
  [arXiv:1902.10548 [hep-ph]].

\bibitem{Ethier:2020way}
  J.~J.~Ethier and E.~R.~Nocera,
  Ann. Rev. Nucl. Part. Sci., no.70, 1-34 (2020)
  [arXiv:2001.07722 [hep-ph]].

\bibitem{Aschenauer:2020pdk}
  E.~C.~Aschenauer, I.~Borsa, G.~Lucero, A.~S.~Nunes and R.~Sassot,
  [arXiv:2007.08300 [hep-ph]].

\bibitem{Gamberg:2006ru}
  L.~P.~Gamberg, D.~S.~Hwang, A.~Metz and M.~Schlegel,
  Phys. Lett. B \textbf{639}, 508-512 (2006)
  [arXiv:hep-ph/0604022 [hep-ph]].

\bibitem{Kovchegov:2018znm}
  Y.~V.~Kovchegov and M.~D.~Sievert,
  Phys. Rev. D \textbf{99}, 054032 (2019)
  [arXiv:1808.09010 [hep-ph]].

\bibitem{Kovchegov:2018zeq}
  Y.~V.~Kovchegov and M.~D.~Sievert,
  Phys. Rev. D \textbf{99}, 054033 (2019)
  [arXiv:1808.10354 [hep-ph]].

\bibitem{Efremov:2002ut}
  A.~V.~Efremov, K.~Goeke and P.~Schweitzer,
  Phys. Rev. D \textbf{67}, 114014 (2003)
  [arXiv:hep-ph/0208124 [hep-ph]].

\bibitem{Courtoy:2014ixa}
  A.~Courtoy,
  [arXiv:1405.7659 [hep-ph]].

\bibitem{Burkardt:2001iy}
  M.~Burkardt and Y.~Koike,
  Nucl. Phys. B \textbf{632}, 311-329 (2002)
  [arXiv:hep-ph/0111343 [hep-ph]].

\bibitem{Efremov:2002qh}
  A.~V.~Efremov and P.~Schweitzer,
  JHEP \textbf{08}, 006 (2003)
  [arXiv:hep-ph/0212044 [hep-ph]].

\bibitem{Aslan:2018tff}
  F.~Aslan and M.~Burkardt,
  Phys. Rev. D \textbf{101}, 016010 (2020)
  [arXiv:1811.00938 [nucl-th]].

\bibitem{Ma:2020kjz}
  J.~P.~Ma and G.~P.~Zhang,
  [arXiv:2003.13920 [hep-ph]].

\bibitem{Bhattacharya:2020jfj}
  S.~Bhattacharya, K.~Cichy, M.~Constantinou, A.~Metz, A.~Scapellato and 
  F.~Steffens,
  [arXiv:2006.12347 [hep-ph]].

\bibitem{Burkardt:2008ps}
  M.~Burkardt,
  Phys. Rev. D \textbf{88}, 114502 (2013)
  [arXiv:0810.3589 [hep-ph]].

\bibitem{Miller:2007ae}
  G.~A.~Miller,
  Phys. Rev. C \textbf{76}, 065209 (2007)
  [arXiv:0708.2297 [nucl-th]].

\bibitem{Barone:2001sp}
  V.~Barone, A.~Drago and P.~G.~Ratcliffe,
  Phys. Rept. \textbf{359}, 1-168 (2002)
  [arXiv:hep-ph/0104283 [hep-ph]].

\end{thebibliography}
\end{document}